\documentclass[journal,10pt]{IEEEtran}
\usepackage{amsfonts}
\usepackage{cite}
\usepackage{graphicx}
\usepackage{float}
\usepackage{multicol}
\usepackage{bm}
\usepackage{upgreek}
\usepackage{color}
\usepackage{amssymb,amsmath}
\usepackage{balance}
\usepackage{booktabs}
\usepackage{stfloats}
\usepackage{algorithm,algorithmic,float}
\usepackage{algorithm}
\usepackage{algorithmic}
\usepackage{float}
\usepackage{lipsum}
\begin{document}

\title{RIS-Aided Cell-Free Massive MIMO Systems: Joint Design of Transmit Beamforming and Phase Shifts}

\author{Si-Nian Jin, \emph{Member, IEEE}, Dian-Wu Yue, \emph{Senior Member, IEEE}, and Ha H. Nguyen, \emph{Senior Member, IEEE}
\thanks{S.-N. Jin and D.-W. Yue are with the College of Information Science and Technology, Dalian Maritime University, Dalian 116026, China (e-mail: jinsinian@dlmu.edu.cn; dwyue@dlmu.edu.cn).

H. H. Nguyen is with the Department of Electrical and Computer Engineering, University of Saskatchewan, Saskatoon, SK S7N 5A9, Canada (e-mail: ha.nguyen@usask.ca).}}

\maketitle

\begin{abstract}
Cell-free massive MIMO and reconfigurable intelligent surface (RIS) have been recognized as two revolutionary technologies in beyond-fifth-generation wireless networks. This paper studies RIS-aided cell-free massive MIMO systems, where multiple RISs are deployed to assist the communication between multiple access points (APs) and multiple users, with either continuous or discrete phase shifts at the RISs. We formulate the max-min fairness problem that maximizes the minimum achievable rate among all users by jointly optimizing the transmit beamforming at active APs and the phase shifts at passive RISs, subject to power constraints at the APs. To address such a challenging problem, we first study the special single-user scenario and propose an algorithm that can transform the optimization problem into semidefinite program (SDP) or integer linear program (ILP) for the cases of continuous and discrete phase shifts, respectively. By solving the resulting SDP and ILP, we first obtain the optimal phase shifts, and then design the optimal transmit beamforming accordingly. To solve the optimization problem for the multi-user scenario and continuous phase shifts at RISs, we extend the single-user algorithm and propose an alternating optimization algorithm, which can first decompose the max-min fairness problem into two subproblems related to transmit beamforming and phase shifts, and then transform the two subproblems into second-order-cone program and SDP, respectively. For the multi-user scenario and discrete phase shifts, the max-min fairness problem is shown to be a mixed-integer non-linear program (MINLP). To tackle it, we design a ZF-based successive refinement algorithm, which can find a suboptimal transmit beamforming and phase shifts by means of alternating optimization. Numerical results show that compared with benchmark schemes of random phase shifts and without using RISs, the proposed algorithms can significantly increase the minimum achievable rate among all users, especially when the number of reflecting elements at each RIS is large. It is also demonstrated that, compared to the case of programming continuous phase shifts at RISs, using 2-bit discrete phase shifts can practically achieve the same performance.
\end{abstract}

\begin{IEEEkeywords}
Cell-free massive MIMO, reconfigurable intelligent surface, max-min fairness, continuous phase shifts, discrete phase shifts, optimization.
\end{IEEEkeywords}

\IEEEpeerreviewmaketitle

\section{Introduction}

As hundreds of millions of devices are recently joining wireless communication networks around the world, requirements of ubiquitous coverage, higher capacity, higher reliability and lower latency have become urgent and pressing for the beyond-the-fifth-generation (B5G) and the future sixth generation (6G) of wireless communication networks \cite{Giordani20}. In order to meet these requirements, cell-free massive MIMO is considered a promising technology \cite{Ngo17,Mai20,Interdonato19,Ngo18,Bjornson20,EBjornson20,Zhang20}. This is because with cell-free massive MIMO, multiple randomly-distributed access points (APs) are deployed in a given coverage area to provide services to multiple scattered users over the same time/frequency resource blocks \cite{Ngo17}. Recent studies show that cell-free massive MIMO outperforms traditional cellular and small-cell systems in various scenarios \cite{Ngo17,Mai20}. By having effective cooperation among distributed APs, cell-free massive MIMO can effectively alleviate interference, and thus increase the system capacity \cite{Interdonato19}. More recent research works on cell-free massive MIMO include optimizing energy efficiency \cite{Ngo18}, scalability \cite{Bjornson20}, and precoding/beamforming \cite{EBjornson20}.

Although cell-free massive MIMO has many advantages, it also has shortcomings. For example, in order to increase the system capacity further, cell-free massive MIMO usually requires to deploy a larger number of active APs, which will bring high cost and power consumption. Fortunately, there is an emerging technology, called reconfigurable intelligent surface (RIS), which is able to complement massive MIMO and provides an energy-saving alternative to improve the system capacity. A RIS is composed of a large number of passive, low-cost, energy-efficient and high-gain metasurface elements, which can shape the radio waves at the electromagnetic level without the need of digital signal processing \cite{Wu18,Wu19,WuQ20,Wu20,EmilBjornson20}.

As a matter of fact, a number of research works have studied RIS-aided cell-free massive MIMO under various setups \cite{Zhou21,Chien21,Zhang21,YZhang21,Huang21,ZZhang21}. By configuring a single RIS on an unmanned aerial vehicle, an aerial RIS (ARIS)-aided cell-free massive MIMO system was examined in \cite{Zhou21}. In that work, the authors proposed an algorithm to maximize the user's achievable rate by optimizing the power allocation and  beamforming vector at each AP, as well as the continuous phase shifts at the AIRS and the placement of the AIRS. For an RIS-aided cell-free massive MIMO system operating over spatially correlated fading channels, reference \cite{Chien21} introduced an aggregated channel estimation approach and derived closed-form expressions of the uplink and downlink (DL) ergodic throughputs in terms of the channel statistics. For RIS-aided cell-free MIMO systems where several base stations and multiple RISs are coordinated to serve multiple users, the authors in \cite{YZhang21} and \cite{Zhang21} developed a hybrid beamforming scheme and proposed an iterative algorithm to maximize the sum achievable rate and the energy efficiency, respectively. In addition, research on optimizing the network capacity of an RIS-aided cell-free network has also been conducted, such as \cite{Huang21,ZZhang21}.

Regarding the above existing works on RIS-aided cell-free massive MIMO systems, it is pointed out that the works in \cite{Zhou21} and \cite{Chien21} only considered the use of a single RIS instead of the more general case of using multiple RISs to assist the APs to communicate with the users. In addition, \cite{Zhou21} and \cite{Chien21} only considered the continuous phase shifts at the RIS, not the discrete phase shifts that are more practical friendly in implementation. Furthermore, both \cite{Zhou21} and \cite{Chien21} only considered the scenario that all users are far away from the APs, and did not consider the standard cell-free scenario where all APs and users are randomly distributed in a coverage area. Finally, most existing research works focused on maximizing the sum rate of a RIS-aided cell-free system \cite{YZhang21,Huang21,ZZhang21} or the single user's rate of an aerial RIS-aided cell-free massive MIMO system \cite{Zhou21}. However, an important feature and expectation of a cell-free system is to ensure that all users in different locations enjoy good services \cite{Ngo17}. In this regard, to the authors' best knowledge, there are no existing works that study the problem of maximizing the minimum achievable rate of all users in an RIS-aided cell-free massive MIMO system by jointly optimizing transmit beamforming and continuous/discrete phase shifts. To address the above mentioned research gap, this paper makes the following specific contributions:
\begin{itemize}
  \item We propose a RIS-aided cell-free massive MIMO architecture by replacing some active APs with some passive and energy-efficient RISs. For such a system,  multiple single-antenna APs provide services to multiple single-antenna users with the assistance of multiple RISs equipped with a large number of scattering elements. In order to guarantee that all users enjoy uniformly good service, we formulate a max-min rate optimization that maximizes the minimum achievable rate among users by jointly optimizing the continuous transmit beamforming at all APs and the continuous/discrete phase shifts at all RISs, subject to given transmit powers at different APs. Due to the coupling of transmit beamforming and phase shifts, the formulated max-min optimization problem is non-convex and challenging to solve.

  \item For the single-user setup and when all RISs are programmed with continuous phase shifts, we propose an algorithm that makes use of semidefinite relaxation (SDR) to transform the max-min fairness problem to a semidefinite program (SDP), which can be effectively solved by a convex optimization toolbox \cite{Boyd04}. With our proposed algorithm, the user's achievable rate can be maximized by first optimizing the continuous phase shifts and then designing the transmit beamforming accordingly. Built on the proposed algorithm for the single-user setup, we extend the optimization framework to the multi-user scenario, and propose an alternating optimization algorithm to solve the max-min fairness problem by designing continuous phase shifts and transmit beamforming in an iterative manner.

  \item For the case that the system uses discrete phase shifts at all RISs and provides services to a single user, we propose an algorithm that can transform the max-min fairness problem into a binary integer linear program (ILP), which can be optimally solved by the branch-and-bound method \cite{Burer12}. By solving the binary ILP, we first obtain the optimal discrete phase shifts, and then design the optimal transmit beamforming based on the obtained phase shifts. On the other hand, for the multi-user setup with discrete phase shifts, the max-min fairness problem is a mixed-integer non-linear program (MINLP), which is difficult to solve optimally. For this case, we propose a zero-forcing (ZF)-based successive refinement algorithm and find the suboptimal transmit beamforming and phase shifts via an alternating optimization method.

  \item Simulation results demonstrate that, compared to benchmark schemes, our proposed algorithms can significantly improve the rate of the user having the worst quality of service (QoS). In addition, for the $95{\%}$-likely minimum achievable rate among all users, the proposed algorithm for the case of discrete phase shifts with 2-bit resolution can basically achieve $95{\%}$ and $85{\%}$ of the performance as compared to the continuous phase shift algorithm for the single-user setup and the multi-user setup, respectively.
\end{itemize}

The remainder of the paper is organized as follows. Section II introduces the system model of RIS-aided cell-free massive MIMO and presents problem formulations for both cases of continuous and discrete phase shifts at IRSs. In Section III, we develop two algorithms to solve the optimization problems for the special single-user system. Section IV extends the algorithms in Section III and proposes an alternating optimization algorithm and a successive refinement algorithm for the more general multi-user system. Simulation results are provided and discussed in Section V to validate the advantages of our proposed algorithms. Finally, Section VI concludes the paper.

\emph{Notations:} The superscripts ${\left( {\cdot} \right)^*}$, ${\left( {\cdot} \right)^{\rm T}}$ and ${\left( {\cdot} \right)^{\rm H}}$ denote conjugate, transpose, conjugate-transpose operations, respectively. $\mathbb{C}^{m\times n}$ denotes the space of $m\times n$ complex-valued matrices. ${\mathbb{E}}\left\{ {\cdot} \right\}$, ${\rm tr}\left( {\cdot} \right)$, $\arg \left( {\cdot} \right)$ and $\left\| {\cdot} \right\|$ represent the statistical expectation, the trace of a matrix, the angle of a complex number, and the Euclidean norm, respectively. ${{\bf{I}}_N}$ denotes an $N \times N$ identity matrix and ${\rm{diag}}\left( {\bf{x}} \right)$ denotes the diagonal matrix whose diagonal elements are in $\bf{x}$. ${\bf{A}} \succeq {\bf{0}}$ indicates a positive semi-definite matrix. ${\left[ {\bf{A}} \right]_{mn}}$ denotes the $\left( {m,n} \right){\mathop{\rm th}\nolimits}$  element of matrix ${\bf{A}}$ and ${\left[ {\bf{a}} \right]_{m}}$ represents the $m$th  element of vector ${\bf{a}}$. The circularly-symmetric complex Gaussian random vector with zero mean and covariance matrix $\bf{A}$ is denoted by $\mathcal{CN}\left( {\bf{0}, \bf{A}} \right)$.

\section{System Model and Problem Formulation}

\subsection{System Model}

\begin{figure}[thb!]
\centering
\includegraphics[scale=0.575]{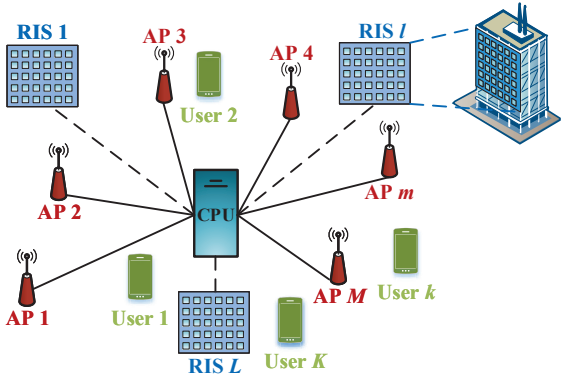}
\caption{A RIS-aided cell-free massive MIMO system.}\label{fig1}
\end{figure}

As illustrated in Fig. \ref{fig1}, we consider a RIS-aided cell-free massive MIMO system, where $M$ single-antenna APs simultaneously serve $K$ single-antenna users with the help of $L$ RISs, each comprising of $N$ scattering elements. As such, the total number of scattering elements at all RISs is $I=LN$. In this system, all APs and RISs are connected to the central processing unit (CPU) through the backhaul link, and follow the instructions of the CPU to provide services to all users on the same time and frequency resource. To better understand the gain brought by the RISs, we assume that all RISs are installed on high-rise buildings so that the incidence rays and reflection rays are not affected by obstacles as much as possible. In addition, in order to represent the actual scene more realistically, we assume that the APs and users may not be in the desired place, and as such, the channels between different APs and users have a certain probability of being blocked by obstacles.

Let $\mathcal{M} = \left\{ {1, \cdots ,M} \right\}$, $\mathcal{K} = \left\{ {1, \cdots ,K} \right\}$, $\mathcal{L} = \left\{ {1, \cdots ,L} \right\}$, $\mathcal{N} = \left\{ {1, \cdots ,N} \right\}$ and $\mathcal{I} = \left\{ {1, \cdots ,I} \right\}$ denote the index sets of the APs, the users, the RISs, the elements at each RIS and the total elements at all RISs, respectively.

\begin{figure}[thb!]
\setlength{\abovecaptionskip}{0.cm}
\setlength{\belowcaptionskip}{-0.cm}
\centering
\includegraphics[scale=0.575]{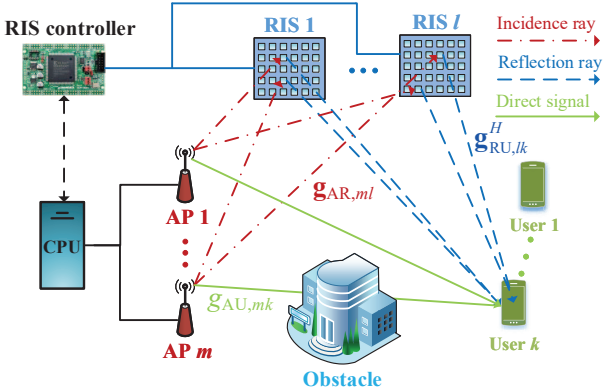}
\caption{Channel model of a RIS-aided cell-free massive MIMO system.}\label{fig2}
\end{figure}
This paper focuses on the DL of a RIS-aided cell-free massive MIMO system and Fig. \ref{fig2} shows the detailed DL channel model. First, all RISs are configured with an RIS controller that coordinates two working modes, namely the receiving mode (channel estimation phase) and the reflecting mode (data transmission phase) \cite{Wu20}. Because of the severe path loss, we only consider the reflected signals that have passed the RISs once, and ignore the reflected signals that have passed through RISs more than once. In addition, we assume the quasi-static flat-fading model for all channels. Finally, in order to characterize the theoretical performance gains brought by the RISs, we assume that the CPU can obtain perfect channel state information (CSI) of all channels, namely the AP-RIS links, RIS-user links, and AP-user links.

As shown in Fig. \ref{fig2}, the notation ${{g}_{{\rm AU},mk}} \in {\mathbb{C}^{1 \times 1}}$ refers to the channel from the $m$th AP to the $k$th user, which is the direct link. We define ${{\bf{g}}_{{\mathop{\rm AU}\nolimits} ,k}} = \left[ {{g_{{\mathop{\rm AU}\nolimits} ,1k}}, \cdots ,{g_{{\mathop{\rm AU}\nolimits} ,Mk}}} \right] \in {\mathbb{C}^{1 \times M}}$ as the channel vector from all APs to the $k$th user. Similarly, the notation ${{\bf{g}}_{{\rm AR},ml}} \in {\mathbb{C}^{N \times 1}}$ denotes the channel from the $m$th AP to the $l$th RIS, which is the incident link, and we define ${{\bf{G}}_{{\mathop{\rm AR}\nolimits} ,l}} = \left[ {{{\bf{g}}_{{\mathop{\rm AR}\nolimits} ,1l}}, \cdots ,{{\bf{g}}_{{\mathop{\rm AR}\nolimits} ,Ml}}} \right] \in {\mathbb{C}^{N \times M}}$ as the channel vector from all APs to the $l$th RIS. The notation ${{\bf{g}}_{{\rm RU},lk}^{\rm H}} \in {\mathbb{C}^{1 \times N}}$ denotes the channel from the $l$th RIS to the $k$th user, which is the reflection link. Each pair of the cascaded channels ${{\bf{g}}_{{\rm AR},ml}}$ and ${{\bf{g}}_{{\rm RU},lk}^{\rm H}}$ from the $m$th AP to the $k$th user via the $l$th RIS can usually be seen as an indirect link which can enhance the communication reliability. To exploit the advantage of the cascaded channel, the phase shifts of the cascaded channels can be adjusted through all RISs. The phase shift matrix at the $l$th RIS can be expressed as the diagonal matrix ${{\bf{\Theta }}_l} = {\mathop{\rm diag}\nolimits} \left( {{{\mathop{\rm e}\nolimits} ^{j{\theta _{l1}}}}, \cdots ,{{\mathop{\rm e}\nolimits} ^{j{\theta _{lN}}}}} \right)$, where ${\theta _{ln}} \in \left[ {0,2\pi } \right)$ denotes the phase shift of the $n$th element at the $l$th RIS. For convenience, we define ${\bf{g}}_{{\mathop{\rm ARU}\nolimits} ,lk} \in {\mathbb{C}^{1 \times M}}$ as the composite AP-RIS-user channel from all APs to the $k$th user via the $l$th RIS, which is given as
\begin{equation}\label{1}
{\bf{g}}_{{\mathop{\rm ARU}\nolimits} ,lk} = {\bf{g}}_{{\mathop{\rm RU}\nolimits} ,lk}^{\rm H}{{\bf{\Theta }}_l}{{\bf{G}}_{{\mathop{\rm AR}\nolimits} ,l}} = {\bf{v}}_l^{\rm H}{{\bf{\Phi }}_{lk}},\;\forall l \in \mathcal{L},\: \forall k \in \mathcal{K},
\end{equation}
where
\begin{equation}\label{2}
{{\bf{v}}_l^{\rm H}}{\rm{ = }}\left[ {{{\rm{e}}^{j{\theta _{l1}}}}, \cdots ,{{\rm{e}}^{j{\theta _{lN}}}}} \right]\in {\mathbb{C}^{1 \times N}},\;\;\;\;\forall l \in \mathcal{L},
\end{equation}
\begin{equation}\label{3}
{{\bf{\Phi }}_{lk}} = {\rm{diag}}\left( {{\bf{g}}_{{\mathop{\rm RU}\nolimits} ,lk}^{\rm H}} \right){{\bf{G}}_{{\mathop{\rm AR}\nolimits} ,l}},\;\;\forall l \in \mathcal{L},\: \forall k \in \mathcal{K}.
\end{equation}

\subsection{Downlink Data Transmission}
During the downlink data transmission phase, the CPU utilizes CSI to precode the intended signal for the $k$th user ($\forall k \in \mathcal{K}$), and then transmits these precoded signals to all users through the $m$th AP ($\forall m \in \mathcal{M}$). Thus, the transmitted signal at the $m$th AP is given as
\begin{equation}\label{4}
{s_m}={{\bf {\bar w}}_m}{\bf{x}}{\rm{ = }}\sum\limits_{k = 1}^K {{w_{mk}}{x_k}},\;\;\;\forall m \in \mathcal{M},
\end{equation}
where ${\bf{x}}{\rm{ = }}\left[ {{x_1}, \cdots ,{x_K}} \right]^{\rm T}$, which satisfies ${\mathop{\rm \mathbb{E}}\nolimits} \left\{ {{\bf{x}}{{\bf{x}}^{\rm H}}} \right\} = {{\bf{I}}_K}$, is the intended signals for $K$ users, and ${{\bf {\bar w}}_m} = \left[ {{w_{m1}}, \cdots ,{w_{mK}}} \right] \in {\mathbb{C}^{1 \times K}}$ denotes the corresponding beamforming vector at the $m$th AP.  In addition, the beamforming vector is chosen to satisfy the following power constraint at each AP \cite{Ngo17}:
\begin{equation}\label{5}
{\left\| {{{{\bf{\bar w}}}_m}} \right\|^2} \le P,\;\;\;\;\forall m \in \mathcal{M}.
\end{equation}
By setting the noise variance to unity, $P$ is also the normalized signal-to-noise ratio (SNR) at the $m$th AP.

With the transmitted signal $s_m$ in \eqref{4}, the received signal at the $k$th user, $r_k$,\:$\forall k \in \mathcal{K}$ is given by
\begin{equation}\label{6}
\begin{array}{l}
{r_k} = \underbrace {\left( {\sum\limits_{l = 1}^L {{\bf{g}}_{{\mathop{\rm RU}\nolimits} ,lk}^{\rm H}{{\bf{\Theta }}_l}{{\bf{G}}_{{\mathop{\rm AR}\nolimits} ,l}}}  + {{\bf{g}}_{{\mathop{\rm AU}\nolimits} ,k}}} \right){{\bf{w}}_k}{x_k}}_{{\rm{desired\: signal}}}\\
 + \underbrace {\sum\limits_{j \ne k}^K {\left( {\sum\limits_{l = 1}^L {{\bf{g}}_{{\mathop{\rm RU}\nolimits} ,lk}^{\rm H}{{\bf{\Theta }}_l}{{\bf{G}}_{{\mathop{\rm AR}\nolimits} ,l}}}  + {{\bf{g}}_{{\mathop{\rm AU}\nolimits} ,k}}} \right){{\bf{w}}_j}{x_j}} }_{{\rm{multiuser\: signal}}} + \underbrace {{n_k}}_{{\rm{noise}}},
\end{array}
\end{equation}
where ${n_k} \sim \mathcal{CN}\left( {0,1} \right)$ denotes the independent and identically distributed (i.i.d.) additive white Gaussian noise (AWGN) at the $k$th user, and ${{\bf{w}}_k} = {\left[ {{w_{1k}}, \cdots ,{w_{Mk}}} \right]^{\rm T}}\in {\mathbb{C}^{M \times 1}}$. Accordingly, the DL signal-to-interference-plus-noise ratio (SINR) and the DL achievable rate of the $k$th user are given by
\begin{equation}\label{8}
{{\mathop{\rm SINR}\nolimits} _k}{\rm{ = }}\frac{{{{\left| {\left( {\sum\limits_{l = 1}^L {{\bf{g}}_{{\mathop{\rm RU}\nolimits} ,lk}^{\rm H}{{\bf{\Theta }}_l}{{\bf{G}}_{{\mathop{\rm AR}\nolimits} ,l}}}  + {{\bf{g}}_{{\mathop{\rm AU}\nolimits} ,k}}} \right){{\bf{w}}_k}} \right|}^2}}}{{\sum\limits_{j \ne k}^K {{{\left| {\left( {\sum\limits_{l = 1}^L {{\bf{g}}_{{\mathop{\rm RU}\nolimits} ,lk}^{\rm H}{{\bf{\Theta }}_l}{{\bf{G}}_{{\mathop{\rm AR}\nolimits} ,l}}}  + {{\bf{g}}_{{\mathop{\rm AU}\nolimits} ,k}}} \right){{\bf{w}}_j}} \right|}^2} + 1} }},
\end{equation}
and
\begin{equation}\label{7}
{R_k} = {\log _2}\left( {1 + {{{\mathop{\rm SINR}\nolimits} }_k}} \right),\;\;\;\forall k \in \mathcal{K}.
\end{equation}

\subsection{Problem Formulation For Continuous Phase Shifts}

In this paper, we show that RIS-aided cell-free massive MIMO systems can ensure uniformly good service to all users, regardless of the users' geographical locations, by jointly optimizing the transmit beamforming $\left\{ {{{\bf{w}}_k}} \right\}$ at all APs and the phase shifts $\left\{{{\bf{\Theta }}_l}\right\}$ at all RISs. Although the joint transmit beamforming and phase shift design is well studied in RIS-aided wireless communication systems \cite{Wu19}, it has not been done for RIS-aided cell-free massive MIMO systems, especially with the objective of maximizing the minimum achievable rate of all users. In this paper, we aim to find the appropriate transmit beamforming and the phase shift design to solve the max-min fairness problem in \eqref{5}. At the optimum point, all users have the same achievable rate.

Given that the user's achievable rate is directly determined by the user's SINR, the max-min fairness problem can be expressed mathematically as
\begin{align}
\left(\bf{P1}\right): \nonumber\\
\mathop {\max }\limits_{{{\bf{w}}_k},{{\bf{\Theta }}_l}}\mathop {\min }\limits_{{\forall k \in \mathcal{K}}}\;&\frac{{{{\left| {\left( {\sum\limits_{l = 1}^L {{\bf{g}}_{{\mathop{\rm RU}\nolimits} ,lk}^{\rm H}{{\bf{\Theta }}_l}{{\bf{G}}_{{\mathop{\rm AR}\nolimits} ,l}}}  + {{\bf{g}}_{{\mathop{\rm AU}\nolimits} ,k}}} \right){{\bf{w}}_k}} \right|}^2}}}{{\sum\limits_{j \ne k}^K {{{\left| {\left( {\sum\limits_{l = 1}^L {{\bf{g}}_{{\mathop{\rm RU}\nolimits} ,lk}^{\rm H}{{\bf{\Theta }}_l}{{\bf{G}}_{{\mathop{\rm AR}\nolimits} ,l}}}  + {{\bf{g}}_{{\mathop{\rm AU}\nolimits} ,k}}} \right){{\bf{w}}_j}} \right|}^2} + 1} }}\label{9}\\
{\rm s.t.}\;\;\,\,\;\;\;\;&{\left\| {{{{\bf{\bar w}}}_m}} \right\|^2} \le P,\;\;\,\,\forall m \in \mathcal{M},\label{10}\\
&0 \le {\theta _{ln}} < 2\pi ,\,\,\,\forall \,l \in \mathcal{L},\:\forall n \in \mathcal{N}\label{11}.
\end{align}

Although the two constraints in \eqref{10} and \eqref{11} in $\bf{P1}$ are both convex, the objective function \eqref{9} in $\bf{P1}$ is non-convex. In addition, the transmit beamforming and phase shifts are coupled in the objective function. This makes $\bf{P1}$ challenging to solve. However, we shall provide two SDP-based optimization algorithms that can solve $\bf{P1}$ for the single-user and multi-user scenarios in Section III-A and Section IV-A, respectively. Simulation results will also show that the two proposed algorithms can achieve very good performance gains.

\subsection{Problem Formulation for Discrete Phase Shifts}

In Section II-C, we examine the situation that the phase shifts at each RIS can be continuously varied in the range of $\left[ {0,2\pi } \right)$. However, in practical implementation, in order to reduce the hardware cost, the RIS usually selects only a limited number of phase shifts \cite{WuQ20}. In other words, the phase of each element of the RIS can only be set to some discrete values. Specifically, for the case of discrete phase shifts, we assume that the phase shifts at all RISs can only be selected from $B$ uniformly distributed values in the interval $\left[ {0,2\pi } \right)$, where $B = {2^b}$ is the number of phase shift levels and $b$ denotes the number of quantization bits.

Let the index set of discrete phase shifts be given as
\begin{equation}\label{12}
\mathcal{F} = \{ 0,\Delta \theta , \cdots ,\left( {B - 1} \right)\Delta \theta \},
\end{equation}
where $\Delta \theta {\rm{ = }}{{2\pi } \mathord{\left/{\vphantom {{2\pi } B}} \right.\kern-\nulldelimiterspace} B}$. Then, subject to the power constraint of \eqref{5} and the discrete phase shifts of \eqref{12}, the optimization problem of maximizing the minimum achievable rate among all users can be expressed as
\begin{align}
\left(\bf{P2}\right): \nonumber\\
\mathop {\max }\limits_{{{\bf{w}}_k},{{\bf{\Theta }}_l}}\mathop {\min }\limits_{{\forall k \in \mathcal{K}}}\:&\frac{{{{\left| {\left( {\sum\limits_{l = 1}^L {{\bf{g}}_{{\mathop{\rm RU}\nolimits} ,lk}^{\rm H}{{\bf{\Theta }}_l}{{\bf{G}}_{{\mathop{\rm AR}\nolimits} ,l}}}  + {{\bf{g}}_{{\mathop{\rm AU}\nolimits} ,k}}} \right){{\bf{w}}_k}} \right|}^2}}}{{\sum\limits_{j \ne k}^K {{{\left| {\left( {\sum\limits_{l = 1}^L {{\bf{g}}_{{\mathop{\rm RU}\nolimits} ,lk}^{\rm H}{{\bf{\Theta }}_l}{{\bf{G}}_{{\mathop{\rm AR}\nolimits} ,l}}}  + {{\bf{g}}_{{\mathop{\rm AU}\nolimits} ,k}}} \right){{\bf{w}}_j}} \right|}^2} + 1} }}\label{13}\\
{\rm s.t.}\;\;\,\,\;\;\;&{\left\| {{{{\bf{\bar w}}}_m}} \right\|^2} \le P,\;\;\,\,\forall m \in \mathcal{M},\label{14}\\
&{\theta _{ln}} \in \mathcal{F} ,\,\,\,\forall \,l \in \mathcal{L},\:\forall n \in \mathcal{N}\label{15}.
\end{align}

Since the objective function of \eqref{13} is non-convex due to the coupling of $\left\{ {{{\bf{w}}_k}} \right\}$ and $\left\{{{\bf{\Theta }}_l}\right\}$ and all variables of constraints in \eqref{15} are discrete, $\bf{P2}$ is a MINLP and hence NP-hard. Generally, there is no standard method to find the global optimal solution of $\bf{P2}$ \cite{So07,Burer12}. A common solution of $\bf{P2}$ is to relax the constraints of discrete phase shift values in \eqref{15} into continuous values, hence converting $\bf{P2}$ back to $\bf{P1}$. By solving $\bf{P1}$, we obtain the optimized continuous phase shifts, and then choose the discrete phase values closest to the continuous phase shifts as a suboptimal solution. However, the quality of such an suboptimal solution may be poor, especially when the number of quantization bits is small. Different from the approximation method, for the single-user and multi-user scenarios, we proposed an ILP-based algorithm and a ZF-based successive refinement algorithm for solving $\bf{P2}$ in Section III-B and Section IV-B, respectively. Simulation results verify the superiority of our proposed algorithms in Section V.

\section{Single-User System}

In this section, we consider the idealistic scenario that there is only one user in the RIS-aided cell-free massive MIMO system, i.e., $K=1$. In this single-user scenario, there is no interference from any other users, and this scenario helps us to draw important insights into the joint transmit beamforming and phase shift design.

\subsection{Continuous Phase Shifts}

For the case of continuous phase shifts, by dropping the user index, $\bf{P1}$ can be simplified as follows:
\begin{align}
\left(\bf{P3}\right):\:\:\mathop {\max }\limits_{{{\bf{w}}},{{\bf{\Theta }}_l}}\;\;\;&{\left| {\left( {\sum\limits_{l = 1}^L {{\bf{g}}_{{\mathop{\rm RU}\nolimits} ,l}^{\rm H}{{\bf{\Theta }}_l}{{\bf{G}}_{{\mathop{\rm AR}\nolimits} ,l}}}  + {{\bf{g}}_{{\mathop{\rm AU}\nolimits} }}} \right){{\bf{w}}}} \right|^2}\label{16}\\
{\rm s.t.}\;\;\,\,\;&{\left| {{w_{m}}} \right|^2} \le P,\;\;\,\,\forall m \in \mathcal{M},\label{17}\\
&0 \le {\theta _{ln}} < 2\pi ,\,\,\,\forall \,l \in \mathcal{L},\:\forall n \in \mathcal{N}\label{18},
\end{align}
where ${\bf{w}} = {\left[ {{w_1}, \cdots ,{w_M}} \right]^{\rm T}}\in {\mathbb{C}^{M \times 1}}$. Although $\bf{P3}$ is simpler than $\bf{P1}$, the objective function of \eqref{16} still has the coupling of ${{\bf{w}}}$ and $\left\{{{\bf{\Theta }}_l}\right\}$, and hence $\bf{P3}$ is still a non-convex optimization problem. In order to solve this problem, we decouple ${{\bf{w}}}$ and $\left\{{{\bf{\Theta }}_l}\right\}$ in the objective function, and then solve the decoupled problem.

For convenience, we define ${{\bf{\bar g}}_{{\mathop{\rm AU}\nolimits} }} = \sum\nolimits_{l = 1}^L {{\bf{g}}_{{\mathop{\rm RU}\nolimits} ,l}^{\rm H}{{\bf{\Theta }}_l}{{\bf{G}}_{{\mathop{\rm AR}\nolimits} ,l}}}  + {{\bf{g}}_{{\mathop{\rm AU}\nolimits} }} \in {\mathbb{C}^{1 \times M}}$, where the $m$th element of ${{\bf{\bar g}}_{{\mathop{\rm AU}\nolimits} }}$ is expressed as ${\bar g_{{\mathop{\rm AU}\nolimits} ,m}}$. For any given ${{\bf{\Theta }}_l},\forall l \in \mathcal{L}$, we observe that in order to maximize the objective function of \eqref{16} the optimal transmit beamforming is the maximum-ratio transmission (MRT) \cite{Tse05}. Therefore, by satisfying the power constraint of \eqref{17}, the optimal transmit beamforming $w_m$ at the $m$th AP is given as ${w_m} = \sqrt P {{\bar g_{{\mathop{\rm AU}\nolimits} ,m}^*} \mathord{\left/{\vphantom {{\bar g_{{\mathop{\rm AU}\nolimits} ,m}^*} {\left| {\bar g_{{\mathop{\rm AU}\nolimits} ,m}^*} \right|}}} \right.\kern-\nulldelimiterspace} {\left| {\bar g_{{\mathop{\rm AU}\nolimits} ,m}} \right|}}$. Substituting this into \eqref{16}, the objective function can be expressed as follows:
\begin{equation}\label{19}
{\left| {\left( {\sum\limits_{l = 1}^L {{\bf{g}}_{{\mathop{\rm RU}\nolimits} ,l}^{\rm H}{{\bf{\Theta }}_l}{{\bf{G}}_{{\mathop{\rm AR}\nolimits} ,l}}}  + {{\bf{g}}_{{\mathop{\rm AU}\nolimits} }}} \right){\bf{w}}} \right|^2} = P{\left( {\sum\limits_{m = 1}^M {\left| {{{\bar g}_{{\mathop{\rm AU}\nolimits} ,m}}} \right|} } \right)^2}.
\end{equation}

Examining the right side of \eqref{19}, we have the following inequality
\begin{equation}\label{20}
\begin{aligned}
&P{\left( {\sum\limits_{m = 1}^M {\left| {{{\bar g}_{{\mathop{\rm AU}\nolimits} ,m}}} \right|} } \right)^2} \mathop  \ge \limits^{\left( {\rm{a}} \right)} P\sum\limits_{m = 1}^M {{{\left| {{{\bar g}_{{\mathop{\rm AU}\nolimits} ,m}}} \right|}^2}} \\
&= P\left( {{{{\bf{ v}}}^{\rm H}}{\bf{\Phi }}{{{\bf{ \Phi }}}^{\rm H}}{\bf{v}} + {{{\bf{ v}}}^{\rm H}}{\bf{ \Phi g}}_{{\mathop{\rm AU}\nolimits} }^{\rm H} + {{\bf{g}}_{{\mathop{\rm AU}\nolimits} }}{{{\bf{ \Phi }}}^{\rm H}}{\bf{ v}} + {{\bf{g}}_{{\mathop{\rm AU}\nolimits} }}{\bf{g}}_{{\mathop{\rm AU}\nolimits} }^{\rm H}} \right),
\end{aligned}
\end{equation}
where ${\bf{v}}_l^{\rm H}$ is given in \eqref{2} and
\begin{equation}\label{21}
{{\bf{v}}^{\rm H}}{\rm{ = }}\left[ {{\bf{v}}_1^{\rm H}, \cdots ,{\bf{v}}_L^{\rm H}} \right] \in {\mathbb{C}^{1 \times I}},
\end{equation}
\begin{equation}\label{22}
{{\bf{\Phi }}_l} = {\rm{diag}}\left( {{\bf{g}}_{{\mathop{\rm RU}\nolimits} ,l}^{\rm H}} \right){{\bf{G}}_{{\mathop{\rm AR}\nolimits} ,l}} \in {\mathbb{C}^{N \times M}},\;\;\forall l \in \mathcal{L},
\end{equation}
\begin{equation}\label{23}
{\bf{\Phi }}{\rm{ = }}{\left[ {{\bf{\Phi }}_1^{\rm T}, \cdots ,{\bf{\Phi }}_L^{\rm T}} \right]^{\rm T}} \in {\mathbb{C}^{I \times M}}.
\end{equation}

The equality of (a) in \eqref{20} holds if and only if $\left| {{{\bar g}_{{\mathop{\rm AU}\nolimits} ,m}}} \right| = 0,\;\forall m \in \mathcal{M}$. Since $P\sum\nolimits_{m = 1}^M {{{\left| {{{\bar g}_{{\mathop{\rm AU}\nolimits} ,m}}} \right|}^2}}$ is a lower bound of $P{\left( {\sum\nolimits_{m = 1}^M {\left| {{{\bar g}_{{\mathop{\rm AU}\nolimits} ,m}}} \right|} } \right)^2}$, it is reasonable to maximize it instead of maximizing the objective function $P{\left( {\sum\nolimits_{m = 1}^M {\left| {{{\bar g}_{{\mathop{\rm AU}\nolimits} ,m}}} \right|} } \right)^2}$. In other words, we can use $P\sum\nolimits_{m = 1}^M {{{\left| {{{\bar g}_{{\mathop{\rm AU}\nolimits} ,m}}} \right|}^2}}$ as the new objective function.

According to \eqref{20} and the fact that ${{\bf{g}}_{{\mathop{\rm AU}\nolimits} }}{\bf{g}}_{{\mathop{\rm AU}\nolimits} }^{\rm H}$ is a constant regardless of ${{\bf{v}}^{\rm H}}$, $(\bf{P3})$ can be reformulated as
\begin{align}
\left( {\bf{P4}} \right):\;\;\mathop {\max }\limits_{{\bf v}}\;\;&{{{\bf{v}}}^{\rm H}}{\bf{\Phi }}{{{\bf{\Phi }}}^{\rm H}}{\bf{v}} + {{{\bf{ v}}}^{\rm H}}{\bf{\Phi g}}_{{\mathop{\rm AU}\nolimits} }^{\rm H} + {{\bf{g}}_{{\mathop{\rm AU}\nolimits} }}{{{\bf{\Phi }}}^{\rm H}}{\bf{v}}\label{24}\\
{\mathop{\rm s.t.}\nolimits} \,\;\;&{\left| {v_i} \right|^2} = 1,\;\;\forall i \in \mathcal{I}\label{25},
\end{align}
where ${\bf{v}}^{\rm H} = {\left[ {{v_1}, \cdots ,{v_I}} \right]} \in {\mathbb{C}^{1 \times I}}$.

We observe that ${\bf{P4}}$ is a non-convex quadratically constrained quadratic program (QCQP). By introducing the auxiliary variable ${\mathord{\buildrel{\lower3pt\hbox{$\scriptscriptstyle\smile$}}\over v} }$, ${\bf{P4}}$ can be equivalently written as the following program of homogeneous QCQP \cite{Luo10}:
\begin{align}
\left( {{\bf{P5}}} \right):\;\mathop {\max }\limits_{{\bf{\tilde v}}}\;\;&{{{\bf{\tilde v}}}^{\rm H}}{\bf{\Omega \tilde v}}\label{26}\\
{\mathop{\rm s.t.}\nolimits} \,\;\;&{\left| {{{\tilde v}_i}} \right|^2} = 1,\;\;\;i = 1, \cdots ,I + 1\label{27},
\end{align}
where
\begin{equation}\label{28}
{{\bf{\tilde v}}^{\rm H}}{\rm{ = }}\left[ {{{\bf{v}}^{\rm H}},\mathord{\buildrel{\lower3pt\hbox{$\scriptscriptstyle\smile$}}
\over v} } \right]= {\left[ {{{\tilde v}_1}, \cdots ,{{\tilde v}_{I+1}}} \right]} \in {\mathbb{C}^{1 \times \left(I + 1\right)}},
\end{equation}
\begin{equation}\label{29}
{\bf{\Omega }}{\rm{ = }}\left[ \begin{array}{l}
{\bf{\Phi }}{{\bf{\Phi }}^{\rm H}}\;\;\;\;\;{\bf{\Phi g}}_{{\mathop{\rm AU}\nolimits} }^{\rm H}\\
{{\bf{g}}_{{\mathop{\rm AU}\nolimits} }}{{\bf{\Phi }}^{\rm H}}\;\;\;\;\;0
\end{array} \right] \in {\mathbb{C}^{\left( {I + 1} \right) \times \left( {I + 1} \right)}}.
\end{equation}

Making use of identities ${{\bf{\tilde v}}^{\rm H}}{\bf{\Omega \tilde v}}={\rm{  tr}}\left( {{\bf{\Omega \tilde V}}} \right)$ and ${\bf{\tilde V}}{\rm{ = }}{\bf{\tilde v}}{{\bf{\tilde v}}^{\rm H}} \in {\mathbb{C}^{\left( {I + 1} \right) \times \left( {I + 1} \right)}}$, $\bf{P5}$ is equivalent to
\begin{align}
\left( {{\bf{P6}}} \right):\;\;\mathop {\max }\limits_{{\bf{\tilde V}}} \;\;&{\mathop{\rm tr}\nolimits} \left( {{\bf{\Omega \tilde V}}} \right)\label{30}\\
{\mathop{\rm s.t.}\nolimits} \,\;\;&{\left[ {{\bf{\tilde V}}} \right]_{ii}} = 1,\;\;\;i = 1, \cdots ,I + 1,\label{31}\\
&{\bf{\tilde V}} \succeq 0,\label{32}\\
&{\mathop{\rm rank}\nolimits} \left( {{\bf{\tilde V}}} \right) = 1\label{33}.
\end{align}

In ${{\bf{P6}}}$, except for the constraint of \eqref{33}, other functions are convex. In order to solve ${{\bf{P6}}}$, we apply the SDR method to relax the constraint of \eqref{33} \cite{Luo10} and reformulate ${{\bf{P6}}}$ as
\begin{align}
\left( {{\bf{P7}}} \right):\;\;\mathop {\max }\limits_{{\bf{\tilde V}}} \;\;&{\mathop{\rm tr}\nolimits} \left( {{\bf{\Omega \tilde V}}} \right)\label{34}\\
{\mathop{\rm s.t.}\nolimits} \,\;\;&\eqref{31},\eqref{32}.\label{35}
\end{align}

Clearly, ${{\bf{P7}}}$ is a standard convex SDP, which can be effectively solved by existing optimization solver (e.g. MOSEK), even when the scale of ${{\bf{\tilde V}}}$ is large. It is pointed out, however, that the rank of the obtained solution is generally not equal to 1 (i.e. ${\mathop{\rm rank}\nolimits} \left( {{\bf{\tilde V}}} \right) \ne 1$). This means that the optimal objective value of ${{\bf{P7}}}$ can only be used as the upper bound of the optimal objective value of ${{\bf{P5}}}$. Therefore, we need to add other steps to reconstruct rank-one solution from the high-rank solution of ${{\bf{P7}}}$. The specific steps are as follows \cite{Wu18}. First, we need to perform eigenvalue decomposition on ${\bf{\tilde V}}{\rm{ = }}{\bf{U}}{\bf \Sigma} {{\bf{U}}^{\rm H}}$, where ${\bf{U}} \in {\mathbb{C}^{\left( {I + 1} \right) \times \left( {I + 1} \right)}}$ and ${\mathbf{\Sigma}}  \in {\mathbb{C}^{\left( {I + 1} \right) \times \left( {I + 1} \right)}}$ represent the decomposed unitary matrix and diagonal matrix, respectively. Next, we construct a suboptimal solution of ${{\bf{P5}}}$ through ${\bf{U}}$ and $\mathbf{\Sigma}$ and this suboptimal solution is given as ${\bf{\tilde v}} = {\bf{U}}{{\bf{\Sigma }}^{{1 \mathord{\left/{\vphantom {1 2}} \right.\kern-\nulldelimiterspace} 2}}}{\bf{\tilde u}} \in {\mathbb{C}^{\left( {I + 1} \right) \times 1}}$, where ${\bf{\tilde u}} \sim \mathcal{CN}\left( {{\bf{0}},{{\bf{I}}_{I + 1}}} \right)$. Then, by generating a large number of  ${\bf{\tilde v}}$ independently, we choose ${\bf{\tilde v}}$ that can maximize the objective value of ${{\bf{P5}}}$ as our final result. Finally, the suboptimal solution ${\bf v}$ of  ${{\bf{P4}}}$ can be recovered by ${\bf{\tilde v}}$, and the $i$th element of $\bf v$ is given by ${\left[ {\bf{v}} \right]_i} = {{\mathop{\rm e}\nolimits} ^{j\arg \left( {{{{{\left[ {{\bf{\tilde v}}} \right]}_i}} \mathord{\left/
{\vphantom {{{{\left[ {{\bf{\tilde v}}} \right]}_i}} {{{\left[ {{\bf{\tilde v}}} \right]}_{_{I + 1}}}}}} \right.
\kern-\nulldelimiterspace} {{{\left[ {{\bf{\tilde v}}} \right]}_{_{I + 1}}}}}} \right)}},\;\forall i \in \mathcal{I}$. According to \cite{Luo10}, when the number of randomly generated ${\bf{\tilde v}}$ is large enough, substituting the suboptimal solution $\bf v$ into \eqref{24} can reach up $\frac{\pi }{4}-$approximation of the optimal objective value of  ${{\bf{P4}}}$.

By the above method, we can find optimized phase shifts $\bf v$ that can maximize the objective value of ${{\bf{P4}}}$. Substituting ${{{\bf{\Theta }}_l}}$, whose $n$th element is ${\left[ {{{\bf{\Theta }}_l}} \right]_{nn}} = \left[ {\bf{v}} \right]_{\left( {l - 1} \right)N + n}^ *$, into ${\bf{\bar g}}_{{\mathop{\rm AU}\nolimits} }$, we can get the optimal transmit beamforming $\bf w$ with the $m$th element being ${w_m} = \sqrt P {{\bar g_{{\mathop{\rm AU}\nolimits} ,m}^*} \mathord{\left/{\vphantom {{\bar g_{{\mathop{\rm AU}\nolimits} ,m}^*} {\left| {\bar g_{{\mathop{\rm AU}\nolimits} ,m}^*} \right|}}} \right.\kern-\nulldelimiterspace} {\left| {\bar g_{{\mathop{\rm AU}\nolimits} ,m}} \right|}}$. Finally, we take $\bf w$ and $\left\{{{\bf{\Theta }}_l}\right\}$ as the optimal solution for maximizing the objective value of ${{\bf{P3}}}$. The specific details in solving ${{\bf{P3}}}$ are summarized in ${\bf{Algorithm\:1}}$.

\begin{algorithm}\label{alg:algo1}
\caption{Joint transmit beamforming and phase shift optimization for ${{\bf{P3}}}$}
\begin{algorithmic}[1]
\STATE {Input the channels ${\bf{g}}_{{\mathop{\rm RU}\nolimits} ,l}^{\rm H}$,${{\bf{G}}_{{\mathop{\rm AR}\nolimits} ,l}}$, ${{\bf{g}}_{{\mathop{\rm AU}\nolimits}}}$ and set ${\bf{\Omega }}$.}
\STATE {Solve SDP in ${{\bf{P7}}}$ and obtain the optimal solution ${\bf{\tilde V}}$.}
\STATE {Perform eigenvalue decomposition on ${\bf{\tilde V}}{\rm{ = }}{\bf{U}}{\bf \Sigma} {{\bf{U}}^{\rm H}}$ and construct 1,000 solutions ${\bf{\tilde v}} = {\bf{U}}{{\bf{\Sigma }}^{{1 \mathord{\left/{\vphantom {1 2}} \right.\kern-\nulldelimiterspace} 2}}}{\bf{\tilde u}}$ with ${\bf{\tilde u}} \sim \mathcal{CN}\left( {{\bf{0}},{{\bf{I}}_{I + 1}}} \right)$.}
\STATE {From 1,000 independently generated ${\bf{\tilde v}}$, choose ${\bf{\tilde v}}$ that maximizes ${{{\bf{\tilde v}}}^{\rm H}}{\bf{\Omega \tilde v}}$ as the final solution}.
\STATE {Restore $\bf v$ according to ${\bf{\tilde v}}$, where the $i$th element of $\bf v$ is given by ${\left[ {\bf{v}} \right]_i} = {{\mathop{\rm e}\nolimits} ^{j\arg \left( {{{{{\left[ {{\bf{\tilde v}}} \right]}_i}} \mathord{\left/
 {\vphantom {{{{\left[ {{\bf{\tilde v}}} \right]}_i}} {{{\left[ {{\bf{\tilde v}}} \right]}_{_{I + 1}}}}}} \right.
 \kern-\nulldelimiterspace} {{{\left[ {{\bf{\tilde v}}} \right]}_{_{I + 1}}}}}} \right)}},\;\forall i \in \mathcal{I}$.}
\STATE {Set the optimal phase shifts at the $l$th RIS in ${\bf{\Theta }}_l$, where the $n$th element is ${\left[ {{{\bf{\Theta }}_l}} \right]_{nn}} = \left[ {\bf{v}} \right]_{\left( {l - 1} \right)N + n}^ * $.}
\STATE {Substituting ${{{\bf{\Theta }}_l}}$ into ${{\bf{\bar g}}_{{\mathop{\rm AU}\nolimits} }} = \sum\nolimits_{l = 1}^L {{\bf{g}}_{{\mathop{\rm RU}\nolimits} ,l}^{\rm H}{{\bf{\Theta }}_l}{{\bf{G}}_{{\mathop{\rm AR}\nolimits} ,l}}}  + {{\bf{g}}_{{\mathop{\rm AU}\nolimits} }}$, we obtain the optimal transmit beamforming $\bf w$, where the $m$th element is ${w_m} = \sqrt P {{\bar g_{{\mathop{\rm AU}\nolimits} ,m}^*} \mathord{\left/{\vphantom {{\bar g_{{\mathop{\rm AU}\nolimits} ,m}^*} {\left| {\bar g_{{\mathop{\rm AU}\nolimits} ,m}^*} \right|}}} \right.\kern-\nulldelimiterspace} {\left| {\bar g_{{\mathop{\rm AU}\nolimits} ,m}} \right|}}$.}
\end{algorithmic}
\end{algorithm}

It is worth noting that the complexity of  ${\bf{Algorithm\:1}}$ is dominated by the complexity of Step 2, in which the SDP of ${{\bf{P7}}}$ can be solved by the interior-point method. According to Theorem 3.12 in \cite{Polik10}, the computational complexity of SDP with $p$ SDP constraints which include a $q \times q$ positive semi-definite matrix can be expressed as $\mathcal{O}\left( {p{q^{3.5}} + {p^2}{q^{2.5}} + {p^3}{q^{0.5}}} \right)$. Hence, the computational complexity of ${\bf{Algorithm\:1}}$ is given by $\mathcal{O}\left( {{{\left( {LN + 1} \right)}^{3.5}} + {{\left( {LN + 1} \right)}^{2.5}} + {{\left( {LN + 1} \right)}^{0.5}}}\right)$, which can be approximated as $\mathcal{O}\left( {{L^{3.5}}{N^{3.5}}} \right)$.

\subsection{Discrete Phase Shifts}

For the case of discrete phase shifts, $\bf{P2}$ is simplified to the following by dropping the user index:
\begin{align}
\left( {{\bf{P8}}} \right)\,:\;\mathop {\max }\limits_{{{\bf{w}}},{{\bf{\Theta }}_l}}\;\;\;&{\left| {\left( {\sum\limits_{l = 1}^L {{\bf{g}}_{{\mathop{\rm RU}\nolimits} ,l}^{\rm H}{{\bf{\Theta }}_l}{{\bf{G}}_{{\mathop{\rm AR}\nolimits} ,l}}}  + {{\bf{g}}_{{\mathop{\rm AU}\nolimits} }}} \right){\bf{w}}} \right|^2}\label{36}\\
{\rm s.t.}\;\;\,\,&{\left| {{w_m}} \right|^2} \le P,\;\;\,\,\forall m \in \mathcal{M},\label{37}\\
&{\theta _{ln}} \in \mathcal{F},\,\,\,\forall \,l \in \mathcal{L},\:\forall n \in \mathcal{N}\label{38},
\end{align}
where ${\bf{w}} = {\left[ {{w_1}, \cdots ,{w_M}} \right]^{\rm T}}\in {\mathbb{C}^{M \times 1}}$ and $\mathcal{F}$ is given in \eqref{12}. Similar to the method used for solving $\bf{P3}$, we know that for any given ${{\bf{\Theta }}_l},\forall l \in \mathcal{L}$ the optimal transmit beamforming to maximize the objective function of \eqref{36} is MRT. Therefore, under the condition of satisfying the power constraint of \eqref{37}, the optimal transmit beamforming at the $m$th AP is given as ${w_m} = \sqrt P {{\bar g_{{\mathop{\rm AU}\nolimits} ,m}^*} \mathord{\left/{\vphantom {{\bar g_{{\mathop{\rm AU}\nolimits} ,m}^*} {\left| {\bar g_{{\mathop{\rm AU}\nolimits} ,m}^*} \right|}}} \right.\kern-\nulldelimiterspace} {\left| {\bar g_{{\mathop{\rm AU}\nolimits} ,m}} \right|}}$, where ${{\bf{\bar g}}_{{\mathop{\rm AU}\nolimits} }} = \sum\nolimits_{l = 1}^L {{\bf{g}}_{{\mathop{\rm RU}\nolimits} ,l}^{\rm H}{{\bf{\Theta }}_l}{{\bf{G}}_{{\mathop{\rm AR}\nolimits} ,l}}}  + {{\bf{g}}_{{\mathop{\rm AU}\nolimits} }}$. By using the inequality in \eqref{20}, the objective function of \eqref{36} can be lower bounded as
\begin{align}
&P{\left( {\sum\limits_{m = 1}^M {\left| {{{\bar g}_{{\mathop{\rm AU}\nolimits} ,m}}} \right|} } \right)^2} \ge P{\left\| {\sum\limits_{l = 1}^L {{\bf{g}}_{{\mathop{\rm RU}\nolimits} ,l}^{\rm H}{{\bf{\Theta }}_l}{{\bf{G}}_{{\mathop{\rm AR}\nolimits} ,l}}}  + {{\bf{g}}_{{\mathop{\rm AU}\nolimits} }}} \right\|^2}\label{39}\\
&\:\:\:\:\:\:\:\:\:\:\:\:\:\:\:\:\:\:\:\:= P\left( {{{\bf{v}}^{\rm H}}{\bf{\Psi v}} + 2{\mathop{\rm Re}\nolimits} \left\{ {{{\bf{v}}^{\rm H}}{\bf{\Xi }}} \right\} + {{\bf{g}}_{{\mathop{\rm AU}\nolimits} }}{\bf{g}}_{{\mathop{\rm AU}\nolimits} }^{\rm H}} \right)\label{40},
\end{align}
where ${\bf{v}}^{\rm H}$ and ${{\bf{\Phi }}}$ are given in \eqref{21} and \eqref{23}, respectively, and other parameters are
\begin{equation}\label{41}
{\bf{\Psi }} = {\bf{\Phi }}{{\bf{\Phi }}^{\rm H}} \in {\mathbb{C}^{I \times I}},
\end{equation}
\begin{equation}\label{42}
{\bf{\Xi }} = {\bf{\Phi g}}_{{\mathop{\rm AU}\nolimits} }^{\rm H} \in {\mathbb{C}^{I \times 1}}.
\end{equation}

Based on \eqref{40}, $\bf{P8}$ can be reformulated as
\begin{align}
\left( {{\bf{P9}}} \right):\;\;\;\mathop {\max }\limits_{{\bm \uptheta} } \;\;&{{\bf{v}}^{\rm H}}{\bf{\Psi v}} + 2{\mathop{\rm Re}\nolimits} \left\{ {{{\bf{v}}^{\rm H}}{\bf{\Xi }}} \right\}\label{43}\\
\;\;\;\;\;\;\;\;\;\;\;\;{\mathop{\rm s.t.}\nolimits} \,\;\;&{\theta _{i}} \in \mathcal{F},\,\,\,\forall \,i \in \mathcal{I} \label{44},
\end{align}
where ${\bm \uptheta} {\rm{ = }}\left[ {{\theta _1}, \cdots ,{\theta _{I}}} \right]$ and ${{\bf{v}}^{\rm H}} = \left[ {{{\mathop{\rm e}\nolimits} ^{j{\theta _1}}}, \cdots ,{{\mathop{\rm e}\nolimits} ^{j{\theta _{I}}}}} \right]\in {\mathbb{C}^{1 \times I}}$. Note that the $i = \left( {l - 1} \right)N + n$th element of ${\bm \uptheta}$ denotes the phase shift of $n$th element at the $l$th RIS.

Although $\bf{P9}$ is still a non-convex problem, its can be reformulated as an ILP. Specifically, we can first utilize the special structure of the objective function to convert $\bf{P9}$ into a binary (zero-one) ILP and then obtain an optimal solution by solving the binary ILP. Before performing the relevant conversions, we first define a special ordered set of type 1 (SOS1). The SOS1 denotes a set of vectors ${\bf{z}}$ with the size of $B \times 1$, in which only one element can be 1 and all other elements are 0. It can be expressed mathematically as \cite{Burer12}
\begin{equation}\label{45}
\sum\limits_{b = 1}^{B} {{{\left[ {\bf{z}} \right]}_{b}}}  = 1,\;\;{\left[ {\bf{z}} \right]_{b}} \in \left[ {0,1} \right].
\end{equation}

Performing some algebra operations and using the fact that $\bf{\Psi}$ is a positive semi-definite matrix, the objective function of \eqref{43} in $\bf{P9}$ is equivalent to
\begin{equation}\label{46}
\begin{array}{l}
f\left( {\bm{\uptheta }},{\bm{\upvarphi }} \right) \buildrel \Delta \over = \sum\limits_{i = 1}^{I} {2\left| {{{\left[ {\bf{\Xi }} \right]}_i}} \right|\cos \left( {\arg \left( {{{\left[ {\bf{\Xi }} \right]}_i}} \right)} \right)\cos \left( {{\theta _i}} \right)} \\
 - \sum\limits_{i = 1}^{I} {2\left| {{{\left[ {\bf{\Xi }} \right]}_i}} \right|\sin \left( {\arg \left( {{{\left[ {\bf{\Xi }} \right]}_i}} \right)} \right)\sin \left( {{\theta _i}} \right)} + \sum\limits_{i = 1}^I {{{\left[ {\bf{\Psi }} \right]}_{ii}}} \\
{\rm{ + }}\sum\limits_{{i_1} = 1}^{I - 1} {\sum\limits_{{i_2} = {i_1} + 1}^{I} {2\left| {{{\left[ {\bf{\Psi }} \right]}_{{i_1}{i_2}}}} \right|\cos \left( {\arg \left( {{{\left[ {\bf{\Psi }} \right]}_{{i_1}{i_2}}}} \right)} \right)\cos \left( {{\varphi _{{i_1},{i_2}}}} \right)} } \\
 - \sum\limits_{{i_1} = 1}^{I - 1} {\sum\limits_{{i_2} = {i_1} + 1}^{I} {2\left| {{{\left[ {\bf{\Psi }} \right]}_{{i_1}{i_2}}}} \right|\sin \left( {\arg \left( {{{\left[ {\bf{\Psi }} \right]}_{{i_1}{i_2}}}} \right)} \right)\sin \left( {{\varphi _{{i_1},{i_2}}}} \right)} },
\end{array}
\end{equation}
where ${\left[ {\bm{\upvarphi }} \right]_{{i_1}{i_2}}} = {\varphi _{{i_1},{i_2}}} = {\theta _{{i_1}}} - {\theta _{{i_2}}},\forall {i_1},{i_2} \in \mathcal{I}\;$. Based on \eqref{46}, $\bf{P9}$ can be recast as
\begin{align}
\left( {{\bf{P10}}} \right):\;\;\mathop {\max }\limits_{{\bm \uptheta ,\bm \upvarphi} }\;\;&f\left( {\bm \uptheta ,\bm \upvarphi } \right)\label{47}\\
{\mathop{\rm s.t.}\nolimits} \,\;\;&{\theta _i} \in \mathcal{F},\,\,\,\forall \,i \in \mathcal{I},\label{48}\\
&{\varphi _{{i_1},{i_2}}} = {\theta _{{i_1}}} - {\theta _{{i_2}}},\;\;\forall {i_1},{i_2} \in \mathcal{I}\label{49}.
\end{align}

We notice that ${\cos \left( {{\theta _i}} \right)}$, ${\sin \left( {{\theta _i}} \right)}$, ${\cos \left( {{\varphi _{{i_1},{i_2}}}} \right)}$ and ${\sin \left( {{\varphi _{{i_1},{i_2}}}} \right)}$ in $f\left( {\bm \uptheta ,\bm \upvarphi } \right)$ are non-linear functions with respect to the associated optimization variables ${\bm \uptheta}$ and ${\bm \upvarphi}$. In order to convert $f\left( {\bm \uptheta ,\bm \upvarphi } \right)$ into a linear function, we first introduce the SOS1 binary vectors ${{\bf{z}}_i},\forall i \in \mathcal{I}$ with the size of $B \times 1$ to convert ${\cos \left( {{\theta _i}} \right)}$ and ${\sin \left( {{\theta _i}} \right)}$ to linear functions with respect to the optimization variable ${\bm \uptheta}$. To this end, we define three vectors ${\bf{y}}_1$, ${\bf{y}}_2$ and ${\bf{y}}_3$ with the size of $B \times 1$ as follows:
\begin{align}
    &{{\bf{y}}_1}{\rm{ = }}\left[ {0,\Delta \theta , \cdots ,\left( {B - 1} \right)\Delta \theta } \right]^{\rm T},\label{50}\\
     &{{\bf{y}}_2}{\rm{ = }}\left[ {\cos \left( 0 \right),\cos \left( {\Delta \theta } \right), \cdots ,\cos \left( {\left( {B - 1} \right)\Delta \theta } \right)} \right]^{\rm T},\label{51}\\
     &{{\bf{y}}_3}{\rm{ = }}\left[ {\sin \left( 0 \right),\sin \left( {\Delta \theta } \right), \cdots ,\sin \left( {\left( {B - 1} \right)\Delta \theta } \right)} \right]^{\rm T}\label{52}.
\end{align}

According to the definitions of ${{\bf{z}}_i}$, ${\bf{y}}_1$, ${\bf{y}}_2$ and ${\bf{y}}_3$, the variables ${\theta _i}$, $\cos \left( {{\theta _i}} \right)$ and $\sin \left( {{\theta _i}} \right)$ appearing in $f\left( {\bm \uptheta ,\bm \upvarphi } \right)$ can be expressed as ${\theta _i} = {\bf{y}}_1^{\rm T}{{\bf{z}}_i}$, $\cos \left( {{\theta _i}} \right){\rm{ = }}{\bf{y}}_2^{\rm T}{{\bf{z}}_i}$ and $\sin \left( {{\theta _i}} \right){\rm{ = }}{\bf{y}}_3^{\rm T}{{\bf{z}}_i}$.

Next, in order to convert ${\cos \left( {{\varphi _{{i_1},{i_2}}}} \right)}$ and ${\sin \left( {{\varphi _{{i_1},{i_2}}}} \right)}$ to linear functions with respect to the associated optimization variables ${\bm \uptheta}$ and ${\bm \upvarphi}$, we introduce the SOS1 binary vectors ${{\bf{\bar z}}_{{i_1},{i_2}}},\forall {i_1},{i_2} \in \mathcal{I}$ with the size of $B \times 1$ and the binary variables ${\tilde z_{{i_1},{i_2}}\in \left\{ {0,1} \right\}} ,\forall {i_1},{i_2} \in \mathcal{I}$. The reason for introducing  ${\tilde z_{{i_1},{i_2}}}$ is because the values of ${\theta _{{i_1}}}$ and ${\theta _{{i_2}}}$ are from $\mathcal{F} = \{ 0,\Delta \theta , \cdots ,\left( {B - 1} \right)\Delta \theta \}$, which makes the range of ${\varphi _{{i_1},{i_2}}} = {\theta _{{i_1}}} - {\theta _{{i_2}}}$ to be $\mathcal{\bar F} = \{  - \left( {B - 1} \right)\Delta \theta , \cdots ,0, \cdots ,\left( {B - 1} \right)\Delta \theta \}$. To change the range of ${\varphi _{{i_1},{i_2}}}$  back to $\mathcal{F} = \{ 0,\Delta \theta , \cdots ,\left( {B - 1} \right)\Delta \theta \}$, we can set the binary variable ${\tilde z_{{i_1},{i_2}}}=0$ when ${\varphi _{{i_1},{i_2}}}{\rm{ = }}{\theta _{{i_1}}} - {\theta _{{i_2}}}$ belongs to $\left[ {0,2\pi } \right)$, and ${\tilde z_{{i_1},{i_2}}}=1$ when ${\varphi _{{i_1},{i_2}}}{\rm{ = }}{\theta _{{i_1}}} - {\theta _{{i_2}}}$ belongs to $\left( { - 2\pi ,0} \right)$.

According to the definitions of ${{\bf{\bar z}}_{{i_1},{i_2}}}$, ${\tilde z_{{i_1},{i_2}}}$, ${\bf{y}}_1$, ${\bf{y}}_2$ and ${\bf{y}}_3$, the variables $\varphi _{{i_1},{i_2}}$, ${\cos \left( {{\varphi _{{i_1},{i_2}}}} \right)}$ and ${\sin \left( {{\varphi _{{i_1},{i_2}}}} \right)}$ appearing in $\bf{P10}$ can be expressed as
\begin{align}
&{\varphi _{{i_1},{i_2}}} =  - 2\pi {{\tilde z} _{{i_1},{i_2}}} + {\bf{y}}_1^{\rm T}{{\bf{\bar z}}_{{i_1},{i_2}}}\label{53},\\
&\cos \left( {{\varphi _{{i_1},{i_2}}}} \right) = {\bf{y}}_2^{\rm T}{{\bf{\bar z}}_{{i_1},{i_2}}}\label{54},\\
&\sin \left( {{\varphi _{{i_1},{i_2}}}} \right) = {\bf{y}}_3^{\rm T}{{\bf{\bar z}}_{{i_1},{i_2}}}\label{55}.
\end{align}

Substituting \eqref{50} to \eqref{55} into $\bf{P10}$, $\bf{P10}$ can be recast as
\begin{align}
\left( {{\bf{P11}}} \right):&\mathop {\max }\limits_{{{\bf{z}}_i},{{{\bf{\bar z}}}_{{i_1},{i_2}}},{{\tilde z}_{{i_1},{i_2}}}}\;\;{\hat f}\left( {{{\bf{z}}_i},{{{\bf{\bar z}}}_{{i_1},{i_2}}}} \right)\label{56}\\
{\mathop{\rm s.t.}\nolimits}\,\;\;\;&{\bf{y}}_1^{\rm T}\left( {{{\bf{z}}_{{i_1}}} - {{\bf{z}}_{{i_2}}}} \right) + 2\pi {{\tilde z}_{{i_1},{i_2}}} = {\bf{y}}_1^{\rm T}{{{\bf{\bar z}}}_{{i_1},{i_2}}},\;\forall {i_1},{i_2} \in \mathcal{I},\label{57}\\
&{{\bf{z}}_{{i}}},{{{\bf{\bar z}}}_{{i_1},{i_2}}} \in {\mathop{\rm SOS}\nolimits} 1,\;\;\;\forall i,{i_1},{i_2} \in \mathcal{I},\label{58}\\
&{{\tilde z}_{{i_1},{i_2}}} \in \left\{ {0,1} \right\},\;\;\;\forall {i_1},{i_2} \in \mathcal{I}\label{59},
\end{align}
where
\begin{equation}\label{60}
\begin{array}{l}
{\hat f}\left( {{{\bf{z}}_{i}},{{\bf{\bar z}}_{i_1,i_2}}} \right) \buildrel \Delta \over = \sum\limits_{i = 1}^{I} {2\left| {{{\left[ {\bf{\Xi }} \right]}_i}} \right|\cos \left( {\arg \left( {{{\left[ {\bf{\Xi }} \right]}_i}} \right)} \right){\bf{y}}_2^{\rm T}{{\bf{z}}_i}} \\
 - \sum\limits_{i = 1}^{I} {2\left| {{{\left[ {\bf{\Xi }} \right]}_i}} \right|\sin \left( {\arg \left( {{{\left[ {\bf{\Xi }} \right]}_i}} \right)} \right){\bf{y}}_3^{\rm T}{{\bf{z}}_i}} + \sum\limits_{i = 1}^I {{{\left[ {\bf{\Psi }} \right]}_{ii}}}\\
{\rm{ + }}\sum\limits_{{i_1} = 1}^{I - 1} {\sum\limits_{{i_2} = {i_1} + 1}^{I} {2\left| {{{\left[ {\bf{\Psi }} \right]}_{{i_1}{i_2}}}} \right|\cos \left( {\arg \left( {{{\left[ {\bf{\Psi }} \right]}_{{i_1}{i_2}}}} \right)} \right){\bf{y}}_2^{\rm T}{{\bf{\bar z}}_{{i_1},{i_2}}}} } \\
 - \sum\limits_{{i_1} = 1}^{I - 1} {\sum\limits_{{i_2} = {i_1} + 1}^{I} {2\left| {{{\left[ {\bf{\Psi }} \right]}_{{i_1}{i_2}}}} \right|\sin \left( {\arg \left( {{{\left[ {\bf{\Psi }} \right]}_{{i_1}{i_2}}}} \right)} \right){\bf{y}}_3^{\rm T}{{\bf{\bar z}}_{{i_1},{i_2}}}} }.
\end{array}
\end{equation}

Since $\bf{P11}$ is a binary ILP, it can be solved quickly by using an existing  optimization solver (e.g. Gurobi). Once we solve $\bf{P11}$, $\bf{P9}$ is also solved accordingly. Therefore, the optimal phase shift of $n$th element at the $l$th RIS can be given as ${\left[ {{{\bf{\Theta }}_l}} \right]_{nn}} = {{\mathop{\rm e}\nolimits} ^{j\left( {{\bf{y}}_1^{\rm T}{{\bf{z}}_i}} \right)}}$, where $i = \left( {l - 1} \right)N + n$. Substituting ${{{\bf{\Theta }}_l}}$ into ${\bf{\bar g}}_{{\mathop{\rm AU}\nolimits} }$, we can obtain the optimal transmit beamforming $\bf w$ with the $m$th element being ${w_m} = \sqrt P {{\bar g_{{\mathop{\rm AU}\nolimits} ,m}^*} \mathord{\left/{\vphantom {{\bar g_{{\mathop{\rm AU}\nolimits} ,m}^*} {\left| {\bar g_{{\mathop{\rm AU}\nolimits} ,m}} \right|}}} \right.\kern-\nulldelimiterspace} {\left| {\bar g_{{\mathop{\rm AU}\nolimits} ,m}} \right|}}$. We then take $\bf w$ and $\left\{{{\bf{\Theta }}_l}\right\}$ as the optimal solution of ${{\bf{P8}}}$. The specific details of solving ${{\bf{P8}}}$ are given in ${\bf{Algorithm\:2}}$.

\begin{algorithm}\label{alg:algo2}
\caption{Joint transmit beamforming and phase shift optimization for ${{\bf{P8}}}$}
\begin{algorithmic}[1]
\STATE {Input the channels ${\bf{g}}_{{\mathop{\rm RU}\nolimits} ,l}^{\rm H}$,${{\bf{G}}_{{\mathop{\rm AR}\nolimits} ,l}}$, ${{\bf{g}}_{{\mathop{\rm AU}\nolimits}}}$, the preset vectors ${\bf{y}}_1$, ${\bf{y}}_2$, ${\bf{y}}_3$, and sets ${\bf{\Psi }}$ and ${\bf{\Xi }}$.}
\STATE {Solve the binary ILP in ${{\bf{P11}}}$ and obtain the optimal solution ${{\bf{z}}_i},\:\forall i \in \mathcal{I}$.}
\STATE {For $i = \left( {l - 1} \right)N + n,\:\forall l \in \mathcal{L},\:\forall n \in \mathcal{N}$, the optimal phase shift of $n$th element at the $l$th RIS is given as ${{\mathop{\rm e}\nolimits} ^{j\left( {{\bf{y}}_1^{\rm T}{{\bf{z}}_i}} \right)}}$, i.e. ${\left[ {{{\bf{\Theta }}_l}} \right]_{nn}} = {{\mathop{\rm e}\nolimits} ^{j\left( {{\bf{y}}_1^{\rm T}{{\bf{z}}_i}} \right)}}$. }
\STATE {Substituting ${{{\bf{\Theta }}_l}}$ into ${{\bf{\bar g}}_{{\mathop{\rm AU}\nolimits} }} = \sum\nolimits_{l = 1}^L {{\bf{g}}_{{\mathop{\rm RU}\nolimits} ,l}^{\rm H}{{\bf{\Theta }}_l}{{\bf{G}}_{{\mathop{\rm AR}\nolimits} ,l}}}  + {{\bf{g}}_{{\mathop{\rm AU}\nolimits} }}$, we can obtain the optimal transmit beamforming $\bf w$ with the $m$th element being ${w_m} = \sqrt P {{\bar g_{{\mathop{\rm AU}\nolimits} ,m}^*} \mathord{\left/{\vphantom {{\bar g_{{\mathop{\rm AU}\nolimits} ,m}^*} {\left| {\bar g_{{\mathop{\rm AU}\nolimits} ,m}} \right|}}} \right.\kern-\nulldelimiterspace} {\left| {\bar g_{{\mathop{\rm AU}\nolimits} ,m}} \right|}}$.}
\end{algorithmic}
\end{algorithm}

In ${\bf{Algorithm\:2}}$, the computational complexity is mainly determined by Step 2, which solves ILP. In order to obtain the optimal solution ${{\bf{z}}_i},\:\forall i \in \mathcal{I}$, ILP can be solved by the branch-and-bound method. Since only one element in ${{\bf{z}}_i}$ is 1, there are at most $B$ possible solutions for ${{\bf{z}}_i}$. Hence, the computational complexity of ${\bf{Algorithm\:2}}$ is $\mathcal{O}\left( B^{LN} \right)$.

\section{Multi-User System}

For the general multi-user system considered in this section, we shall propose two algorithms that can maximize the minimum achievable rate of all users by jointly optimizing the transmit beamforming and phase shifts. One algorithm is for the case of continuous phase shifts, and the other algorithm is for discrete phase shifts.

\subsection{Continuous Phase Shifts}

For this case, the max-min fairness problem has been summarized in $\bf{P1}$. Due to the coupling of the transmit beamforming $\left\{ {{{\bf{w}}_k}} \right\}$ and the phase shift $\left\{{{\bf{\Theta }}_l}\right\}$ in the objective function, $\bf{P1}$ is difficult to solve. Here, we shall propose an alternating optimization algorithm to solve it. First, for any given phase shift ${\bf{\Theta }}_l,\forall l \in \mathcal{L}$, we define the combined channel from all APs to the $k$th user as
\begin{equation}\label{61}
{\bf{\bar g}}_{{\mathop{\rm AU}\nolimits} ,k}^{\rm H} = \sum\limits_{l = 1}^L {{\bf{g}}_{{\mathop{\rm RU}\nolimits} ,lk}^{\rm H}{{\bf{\Theta }}_l}{{\bf{G}}_{{\mathop{\rm AR}\nolimits} ,l}}}  + {{\bf{g}}_{{\mathop{\rm AU}\nolimits} ,k}} \in {\mathbb{C}^{1 \times M}},\forall k \in \mathcal{K}.
\end{equation}

Based on \eqref{61}, $\bf{P1}$ can be written as
\begin{align}
\left( {{\bf{P12}}} \right):\,\mathop {\max }\limits_{{{\bf{w}}_k}}\mathop {\min }\limits_{{\forall k \in \mathcal{K}}}\;\;&\frac{{{{\left| {{\bf{\bar g}}_{{\mathop{\rm AU}\nolimits} ,k}^{\rm H}{{\bf{w}}_k}} \right|}^2}}}{{\sum\limits_{j \ne k}^K {{{\left| {{\bf{\bar g}}_{{\mathop{\rm AU}\nolimits} ,k}^{\rm H}{{\bf{w}}_j}} \right|}^2} + 1} }}\label{62}\\
{\rm s.t.}\;\;\,\,\;\;\;\;&{\left\| {{{{\bf{\bar w}}}_m}} \right\|^2} \le P,\;\;\,\,\forall m \in \mathcal{M}\label{63}.
\end{align}

By introducing variable $\xi$, we reformulate $\bf{P12}$ as follows:
\begin{align}
&\left( {{\bf{P13}}} \right):\;\mathop {\max }\limits_{{{\bf{w}}_k}} \;\;\xi \label{64}\\
&\;\;\;\;{\rm s.t.}\;\;\,{\left| {{\bf{\bar g}}_{{\mathop{\rm AU}\nolimits} ,k}^{\rm H}{{\bf{w}}_k}} \right|^2} \ge \xi\sum\limits_{j \ne k}^K {{{\left| {{\bf{\bar g}}_{{\mathop{\rm AU}\nolimits} ,k}^{\rm H}{{\bf{w}}_j}} \right|}^2} + \xi} ,\;\forall k \in \mathcal{K},\label{65}\\
&\;\;\;\;\;\;\;\;\;\;\;\;{\left\| {{{{\bf{\bar w}}}_m}} \right\|^2} \le P,\;\;\,\,\forall m \in \mathcal{M}\label{66}.
\end{align}

According to \cite{Wiesel06} and \cite{Luo06}, we know that an arbitrary phase rotation can be added to the transmit beamforming vector ${{\bf{w}}_k}$ without affecting the establishment of constraint \eqref{65}. Thus, ${\bf{\bar g}}_{{\mathop{\rm AU}\nolimits} ,k}^{\rm H}{{\bf{w}}_k}$ can be assumed to be real without loss of generality. Based on this fact, $\bf{P13}$ can be rewritten as
\begin{align}
&\left( {{\bf{P14}}} \right):\;\mathop {\max }\limits_{{{\bf{w}}_k}} \;\;\xi \label{67}\\
&\;\;\;\;{\rm s.t.}\;\;\;{\bf{\bar g}}_{{\mathop{\rm AU}\nolimits} ,k}^{\rm H}{{\bf{w}}_k} \ge \sqrt \xi  \left\| {{\bf{\bar g}}_{{\mathop{\rm AU}\nolimits} ,k}^{\rm H}{{\bf{W}}_k},1} \right\|,\forall k \in \mathcal{K},\label{68}\\
&\;\;\;\;\;\;\;\;\;\;\;\;{\left\| {{{{\bf{\bar w}}}_m}} \right\|} \le {\sqrt P},\;\;\,\,\forall m \in \mathcal{M}\label{69},
\end{align}
where ${{\bf{W}}_k}{\rm{ = }}\left[ {{{\bf{w}}_1}, \cdots ,{{\bf{w}}_{k - 1}},{{\bf{w}}_{k + 1}}, \cdots ,{{\bf{w}}_K}} \right] \in {\mathbb{C}^{M \times \left( {K - 1} \right)}}$.

When $\xi$ is assumed to be a constant value, the constraint of \eqref{68} can be regarded as a second-order cone constraint. In addition, the constraint of \eqref{69} is also a second-order cone constraint. Consequently, $\bf{P14}$ can be solved efficiently by a bisection search, where each step solves a feasibility problem of second-order cone programming (SCOP) \cite{Boyd04}. Once we find the optimal transmit beamforming by bisection search, we naturally take it as a final solution of $\bf{P12}$. The specific algorithm for solving $\bf{P12}$ is summarized in ${\bf{Algorithm\:3}}$.
\begin{algorithm}\label{alg:algo3}
\caption{Transmit beamforming optimization for ${{\bf{P12}}}$}
\begin{algorithmic}[1]
\STATE {First, input the channels ${\bf{g}}_{{\mathop{\rm RU}\nolimits} ,lk}^{\rm H}$,${{\bf{G}}_{{\mathop{\rm AR}\nolimits} ,l}}$, ${{\bf{g}}_{{\mathop{\rm AU}\nolimits} ,k}}$, the given phase shift ${{\bf{\Theta }}_l}$, and set ${\bf{\bar g}}_{{\mathop{\rm AU}\nolimits} ,k}^{\rm H},\forall k \in \mathcal{K}$. Then, choose the initial values of ${\xi _{\min }}$ and ${\xi _{\max }}$, which define a range of relevant values of  the objective function in \eqref{67}. Also, set a tolerance value $\varepsilon=0.1$.}
\STATE {Set $\xi {\rm{ = }}{{\left( {{\xi _{\min }} + {\xi _{\max }}} \right)} \mathord{\left/{\vphantom {{\left( {{\xi _{\min }} + {\xi _{\max }}} \right)} 2}} \right.\kern-\nulldelimiterspace} 2}$ and solve the following convex feasibility program:
\begin{equation}\label{70}
\begin{aligned}
{\mathop{\rm Find}\nolimits} \;\;\;&{{\bf{w}}_k},\:\:\:\forall k \in \mathcal{K}\\
{\mathop{\rm s.t.}\nolimits}\:\:\;\;&\eqref{68},\eqref{69},
\end{aligned}
\end{equation}

where ${{\bf{W}}_k}{\rm{ = }}\left[ {{{\bf{w}}_1}, \cdots ,{{\bf{w}}_{k - 1}},{{\bf{w}}_{k + 1}}, \cdots ,{{\bf{w}}_K}} \right]$.}
\STATE {If \eqref{70} is feasible, set ${\xi _{\min }}=\xi$ and set ${\bf{w}}_k,\forall k \in \mathcal{K}$ as the optimial transmit beamforming. Otherwise, set ${\xi _{\max}}=\xi$.}
\STATE {If ${\xi _{\max }} - {\xi _{\min }} \le \varepsilon$, terminate iteration and output the transmit beamforming ${\bf{w}}_k,\forall k \in \mathcal{K}$. Else, go to Step 2.}
\end{algorithmic}
\end{algorithm}

The complexity of  ${\bf{Algorithm\:3}}$ is dominated by solving SCOP in \eqref{70}, in which SCOP can be solved by the interior-point method. According to Theorem 3.11 in \cite{Polik10}, the computational complexity of solving a SCOP problem is $\mathcal{O}\left( {\sqrt r \left( {{p^3} + {p^2}\sum\nolimits_{i = 1}^r {{q_i}}  + \sum\nolimits_{i = 1}^r {q_i^2} } \right)} \right)$, where $p$, $r$, $q_i$ denote the number of optimization variables, the number of SCOP constraints, and the dimension of the $i$th SCOP, respectively. Therefore, the computational complexity of ${\bf{Algorithm\:3}}$ is $\mathcal{O}\left( {{I_{{\mathop{\rm SCOP}\limits} }}\sqrt {M + K} \left( {2{M^3}{K^3} + {M^2}{K^4} + {K^3} + M{K^2}} \right)} \right)$, which can be approximated as  $\mathcal{O}\left( {2{I_{{\mathop{\rm SCOP}\nolimits} }}{M^{3.5}}{K^3}} \right)$, where ${{I_{{\mathop{\rm SCOP}\nolimits} }}}$ denotes the number of iterations for the convergence of ${\bf{Algorithm\:3}}$.

By using ${\bf{Algorithm\:3}}$, we can obtain the optimal transmit beamforming under any given phase shift. On the other hand, for any given transmit beamforming ${\bf{w}}_k,\forall k \in \mathcal{K}$ obtained from ${\bf{Algorithm\:3}}$, $\bf{P1}$ can be reduced to
\begin{align}
\left( {{\bf{P15}}} \right):\;\mathop {\max }\limits_{\bf{v}} \;\mathop {\min }\limits_{\forall k \in \mathcal{K}} \;\;&\frac{{{{\left| {{{\bf{v}}^{\rm H}}{{\bf{u}}_{k,k}} + {{\bar u}_{k,k}}} \right|}^2}}}{{\sum\limits_{j \ne k}^K {\left| {{{\bf{v}}^{\rm H}}{{\bf{u}}_{k,j}} + {{\bar u}_{k,j}}} \right|^2 + 1} }}\label{71}\\
{\rm s.t.}\,\,\;\;\;\;\;\;\;&{\left| {v_i} \right|^2} = 1,\;\;\forall i \in \mathcal{I}\label{72},
\end{align}
where ${\bf{v}}^{\rm H}= {\left[ {{v_1}, \cdots ,{v_I}} \right]}\in {\mathbb{C}^{1 \times I}}$, ${{\bf{\Phi }}_{lk}}$ is given in \eqref{3}, and other parameters are
\begin{equation}\label{73}
{{\bf{\bar \Phi }}_k}{\rm{ = }}{\left[ {{\bf{\Phi }}_{1k}^{\rm T}, \cdots ,{\bf{\Phi }}_{Lk}^{\rm T}} \right]^{\rm T}} \in {\mathbb{C}^{I \times M}},\;\;\forall k \in \mathcal{K},
\end{equation}
\begin{equation}\label{74}
{{\bf{u}}_{k,j}} = {{\bf{\bar \Phi }}_k}{\bf{w}}_j \in {\mathbb{C}^{I \times 1}},\;\;\forall k,j \in \mathcal{K},
\end{equation}
\begin{equation}\label{75}
{\bar u_{k,j}} = {{\bf{g}}_{{\mathop{\rm AU}\nolimits} ,k}}{{\bf{w}}_j},\;\;\forall k,j \in \mathcal{K}.
\end{equation}

By introducing variable $\bar \xi $, we can reformulate $\bf{P15}$ as follows:
\begin{align}
\left( {{\bf{P16}}} \right):\;\mathop {\max }\limits_{\bf{v}}  \;\;&\bar \xi \label{76}\\
{\rm s.t.}\;\;\;&{\left| {{{\bf{v}}^{\rm H}}{{\bf{u}}_{k,k}} + {{\bar u}_{k,k}}} \right|^2} \ge \nonumber\\
&\;\;\bar \xi \sum\limits_{j \ne k}^K {\left| {{{\bf{v}}^{\rm H}}{{\bf{u}}_{k,j}} + {{\bar u}_{k,j}}} \right|^2}  + \bar \xi ,\,\,\,\forall \,k \in \mathcal{K},\label{77}\\
&\;{\left| {{v_i}} \right|^2} = 1,\;\;\;\forall \,i \in \mathcal{I}\label{78}.
\end{align}

Using a method similar to converting $\bf{P4}$ to $\bf{P5}$ and by introducing an auxiliary variable ${\mathord{\buildrel{\lower3pt\hbox{$\scriptscriptstyle\smile$}}\over v} }$, ${\bf{P16}}$ can be equivalently written as:
\begin{align}
\left( {{\bf{P17}}} \right):\;\mathop {\max }\limits_{{\bf{\tilde v}}} &\;\;\bar \xi \label{79}\\
{\rm s.t.}\;&{{{\bf{\tilde v}}}^{\rm H}}{{\bf{\Omega }}_{k,k}}{\bf{\tilde v}} + {\left| {{{\bar u}_{k,k}}} \right|^2} \ge \nonumber\\
\bar \xi &\left( {\sum\limits_{j \ne k}^K {{{{\bf{\tilde v}}}^{\rm H}}{{\bf{\Omega }}_{k,j}}{\bf{\tilde v}}}  + \sum\limits_{j \ne k}^K {{{\left| {{{\bar u}_{k,j}}} \right|}^2}}  + 1} \right),\,\,\forall \,k \in \mathcal{K},\label{80}\\
&{\left| {{{\tilde v}_i}} \right|^2} = 1,\;i = 1, \cdots ,I + 1\label{81},
\end{align}
where
\begin{equation}\label{82}
{{\bf{\tilde v}}^{\rm H}}{\rm{ = }}\left[ {{{\bf{v}}^{\rm H}},\mathord{\buildrel{\lower3pt\hbox{$\scriptscriptstyle\smile$}}
\over v} } \right]= {\left[ {{{\tilde v}_1}, \cdots ,{{\tilde v}_{I+1}}} \right]}  \in {\mathbb{C}^{1 \times \left(I + 1\right)}},
\end{equation}
\begin{equation}\label{83}
{{\bf{\Omega }}_{k,j}} = \left[ \begin{array}{l}
{{\bf{u}}_{k,j}}{\bf{u}}_{k,j}^{\rm H}\;\;{{\bf{u}}_{k,j}}\bar u_{k,j}^*\\
{{\bar u}_{k,j}}{\bf{u}}_{k,j}^{\rm H}\;\;\;\;\;0
\end{array} \right] \in {\mathbb{C}^{\left( {I + 1} \right) \times \left( {I + 1} \right)}},\;\forall k,j \in \mathcal{K}.
\end{equation}

Using properties of the matrix trace operation, we have ${{\bf{\tilde v}}^{\rm H}}{{\bf{\Omega }}_{k,j}}{\bf{\tilde v}} = {\mathop{\rm tr}\nolimits} \left( {{{\bf{\Omega }}_{k,j}}{\bf{\tilde V}}} \right)$, where ${\bf{\tilde V}}{\rm{ = }}{\bf{\tilde v}}{{\bf{\tilde v}}^{\rm H}} \in {\mathbb{C}^{\left( {I + 1} \right) \times \left( {I + 1} \right)}}$. Thus, $\bf{P17}$ can be rewritten as
\begin{align}
\left( {{\bf{P18}}} \right)&:\;\;\mathop {\max }\limits_{{\bf{\tilde V}}} \;\;\bar \xi \label{84}\\
&{\rm s.t.}\;\;{\rm{tr}}\left( {{{\bf{\Omega }}_{k,k}}{\bf{\tilde V}}} \right) + {\left| {{{\bar u}_{k,k}}} \right|^2} \ge \nonumber\\
&\:\:\:\:\bar \xi \left( {\sum\limits_{j \ne k}^K {{\rm{tr}}\left( {{{\bf{\Omega }}_{k,j}}{\bf{\tilde V}}} \right)}  + \sum\limits_{j \ne k}^K {{{\left| {{{\bar u}_{k,j}}} \right|}^2}}  + 1} \right),\,\,\,\forall \,k \in \mathcal{K},\label{85}\\
&\:\:\:\:{\left[ {{\bf{\tilde V}}} \right]_{ii}} = 1,\;\;i = 1, \cdots ,I + 1,\label{86}\\
&\:\:\:\:{\bf{\tilde V}} \succeq 0,\label{87}\\
&\:\:\:\:{\mathop{\rm rank}\nolimits} \left( {{\bf{\tilde V}}} \right) = 1\label{88}.
\end{align}

When $\bar \xi$ is assumed to be a constant value, \eqref{85} can be regarded as a convex SDP constraint. In addition, the constraints of \eqref{86} and \eqref{87} are also convex. However, the rank-one constraint of \eqref{88} is non-convex. In order to solve $\bf{P18}$, we first relax the constraint of \eqref{88} by SDR and transform $\bf{P18}$ to
\begin{align}
\left( {{\bf{P19}}} \right):\;\;\mathop {\max }\limits_{{\bf{\tilde V}}} &\;\;\;\bar \xi  \label{89}\\
{\rm s.t.}&\;\;\eqref{85},\eqref{86},\eqref{87}.\label{90}
\end{align}
It can be seen that $\bf{P19}$ can be solved efficiently by a bisection search, where each step solves a convex feasibility problem \cite{Boyd04}. After solving $\bf{P19}$, we can use a similar Gaussian randomization in ${\bf{Algorithm\:1}}$ to find the phase shift that maximize the objective function of $\bf{P15}$. The specific algorithm for solving $\bf{P15}$ is summarized in ${\bf{Algorithm\:4}}$.

\begin{algorithm}\label{alg:algo4}
\caption{Phase shift optimization for ${{\bf{P15}}}$}
\begin{algorithmic}[1]
\STATE {First, input the channels ${\bf{g}}_{{\mathop{\rm RU}\nolimits} ,lk}^{\rm H}$,${{\bf{G}}_{{\mathop{\rm AR}\nolimits} ,l}}$, ${{\bf{g}}_{{\mathop{\rm AU}\nolimits} ,k}}$, the given transmit beamforming ${\bf{w}}_k$, and set ${{\bf{\Omega }}_{k,j}}$ and ${\bar u_{k,j}}$ for $\forall k,j \in \mathcal{K}$. Then, choose the initial values of ${{\bar \xi} _{\min }}$ and ${{\bar \xi} _{\max }}$, which define a range of relevant values of the objective function in \eqref{89}. Also, set a tolerance value $\varepsilon=0.1$.}
\STATE {Set ${\bar \xi} {\rm{ = }}{{\left( {{{\bar \xi} _{\min }} + {{\bar \xi} _{\max }}} \right)} \mathord{\left/{\vphantom {{\left( {{\xi _{\min }} + {\xi _{\max }}} \right)} 2}} \right.\kern-\nulldelimiterspace} 2}$, and solve the following convex feasibility program:
\begin{equation}\label{91}
\begin{aligned}
{\mathop{\rm Find}\nolimits} \;\;\;&{\bf{\tilde V}}\\
{\mathop{\rm s.t.}\nolimits} \;\;\;\;&\eqref{85},\eqref{86},\eqref{87}.
\end{aligned}
\end{equation}}
\STATE {If \eqref{91} is feasible, set ${{\bar \xi} _{\min }}={\bar \xi}$ and set ${\bf{\tilde V}}$ as the optimal solution. Otherwise, set ${{\bar \xi} _{\max}}={\bar \xi}$.}
\STATE {If ${{\bar \xi} _{\max }} - {{\bar \xi} _{\min }} \le \varepsilon$, output ${\bf{\tilde V}}$ and go to Step 5. Else, go to Step 2.}
\STATE {Perform eigenvalue decomposition on ${\bf{\tilde V}}{\rm{ = }}{\bf{U}}{\bf \Sigma} {{\bf{U}}^{\rm H}}$ and construct 1,000 suboptimal solutions ${\bf{\tilde v}} = {\bf{U}}{{\bf{\Sigma }}^{{1 \mathord{\left/{\vphantom {1 2}} \right.\kern-\nulldelimiterspace} 2}}}{\bf{\tilde u}}$ with ${\bf{\tilde u}} \sim \mathcal{CN}\left( {{\bf{0}},{{\bf{I}}_{I + 1}}} \right)$.}
\STATE {From 1,000 independently generated ${\bf{\tilde v}}$, choose ${\bf{\tilde v}}$ that maximizes the minimum of $\frac{{{{{\bf{\tilde v}}}^{\rm H}}{{\bf{\Omega }}_{k,k}}{\bf{\tilde v}} + {{\left| {{{\bar u}_{k,k}}} \right|}^2}}}{{ {\sum\nolimits_{j \ne k}^K {{{{\bf{\tilde v}}}^{\rm H}}{{\bf{\Omega }}_{k,j}}{\bf{\tilde v}}}  + \sum\nolimits_{j \ne k}^K {{{\left| {{{\bar u}_{k,j}}} \right|}^2}}  + 1} }},\:\forall \,k \in \mathcal{K}$ as a final optimal solution}.
\STATE {Restore $\bf v$ according to ${\bf{\tilde v}}$, where the $i$th element of $\bf v$ is ${\left[ {\bf{v}} \right]_i} = {{\mathop{\rm e}\nolimits} ^{j\arg \left( {{{{{\left[ {{\bf{\tilde v}}} \right]}_i}} \mathord{\left/
 {\vphantom {{{{\left[ {{\bf{\tilde v}}} \right]}_i}} {{{\left[ {{\bf{\tilde v}}} \right]}_{_{I + 1}}}}}} \right.
 \kern-\nulldelimiterspace} {{{\left[ {{\bf{\tilde v}}} \right]}_{_{I + 1}}}}}} \right)}},\;\forall i \in \mathcal{I}$.}
\STATE {Set the optimal phase shifts at the $l$th RIS in ${\bf{\Theta }}_l$, where the $n$th element is ${\left[ {{{\bf{\Theta }}_l}} \right]_{nn}} = \left[ {\bf{v}} \right]_{\left( {l - 1} \right)N + n}^ * $.}
\end{algorithmic}
\end{algorithm}

Similar to the computational complexity of ${\bf{Algorithm\:1}}$, the computational complexity of ${\bf{Algorithm\:4}}$ is $\mathcal{O}\left( {{I_{{\mathop{\rm SDP}\nolimits} }}\left( {p{q^{3.5}} + {p^2}{q^{2.5}} + {p^3}{q^{0.5}}} \right)} \right)$, which can be approximated as $\mathcal{O}\left( {{I_{{\mathop{\rm SDP}\nolimits} }}{K{L^{3.5}}{N^{3.5}}} } \right)$, where $p = K + 1$, $q = LN + 1$ and ${{I_{{\mathop{\rm SDP}\nolimits} }}}$ denotes the number of iterations for the convergence of ${\bf{Algorithm\:4}}$ \cite{Polik10}.

Based on ${\bf{Algorithm\:3}}$ and ${\bf{Algorithm\:4}}$, we propose an alternating optimization algorithm that can solve $\bf{P1}$ and it is summarized in ${\bf{Algorithm\:5}}$.

\begin{algorithm}\label{alg:algo5}
\caption{Alternating optimization algorithm for $\bf{P1}$}
\begin{algorithmic}[1]
\STATE {Input the initial iteration number $t=1$ and the initial phase shift ${{\bf{\Theta }}_l^{(t)}},\forall l \in \mathcal{L}$. Also, set the maximum number of iterations $T=30$.}
\STATE {For the given phase shift ${{\bf{\Theta }}_l^{(t)}}$, solve ${{\bf{P12}}}$ using ${\bf{Algorithm\:3}}$. Then, output the optimal transmit beamforming ${\bf{w}}_k^{(t)},\forall k \in \mathcal{K}$.}
\STATE {For the given transmit beamforming  ${\bf{w}}_k^{(t)}$, solve ${{\bf{P15}}}$ using ${\bf{Algorithm\:4}}$. Then, output the optimal phase shift ${{\bf{\Theta }}_l^{(t+1)}},\forall l \in \mathcal{L}$.}
\STATE {Substitute ${\bf{w}}_k^{(t)},\forall k \in \mathcal{K}$ and ${{\bf{\Theta }}_l^{(t+1)}},\forall l \in \mathcal{L}$ into the objective function of \eqref{9}. Denote the minimum SINR of all users by ${\rm SINR}_{\rm min}^{(t)}$.}
\STATE {If ${\rm SINR}_{\rm min}^{(t)} \leq {\rm SINR}_{\rm min}^{(t-1)}$ or $t=T$, the iteration process is terminated. Otherwise, set $t=t+1$ and go to Step 2.}
\end{algorithmic}
\end{algorithm}

It is worth noticing that when ${\rm SINR}_{\rm min}^{(t)} \leq {\rm SINR}_{\rm min}^{(t-1)}$, ${\bf{Algorithm\:5}}$ will be terminated. This is because the performance of ${\bf{Algorithm\:4}}$ is closely related to the performance of Gaussian randomization. Especially, when the rank of ${\bf{\tilde V}}$ obtained by solving \eqref{91} is larger than one, the phase shift constructed by the Gaussian randomization approach may degrade the achievable rate, hence the alternating optimization of ${\bf{Algorithm\:5}}$ needs to be terminated in this case. Besides, according to the approximate computational complexities of ${\bf{Algorithm\:3}}$ and ${\bf{Algorithm\:4}}$, the approximate computational complexity of ${\bf{Algorithm\:5}}$ can be given as $\mathcal{O}\left( {{I_{{\rm{iter}}}}\left( {{I_{{\mathop{\rm SDP}\nolimits} }}K{L^{3.5}}{N^{3.5}} + 2{I_{{\mathop{\rm SCOP}\nolimits} }}{M^{3.5}}{K^3}} \right)} \right)$, where $I_{{\rm{iter}}}$ denotes the number of iterations for the convergence of ${\bf{Algorithm\:5}}$.

\subsection{Discrete Phase Shifts}

For this case, the optimization problem of maximizing the minimum achievable rate among all users has been formulated in $\bf{P2}$. Due to the coupling of the continuous transmit beamforming $\left\{ {{{\bf{w}}_k}} \right\}$ and the discrete phase shift $\left\{{{\bf{\Theta }}_l}\right\}$ in the objective function, $\bf{P2}$ is a MINLP, which is usually difficult to solve. In this section, we propose a ZF-based successive refinement algorithm to solve it and this algorithm is detailed in the following.

For $\bf{P2}$, the objective function in \eqref{13} reaches the optimal value when the SINRs of all users are the same. In order to enable each user to achieve the same SINR, the suboptimal ZF-based beamforming can be used at all APs. For any given discrete phase ${\theta _{ln}}\in \mathcal{F},\forall \,l \in \mathcal{L},\forall n \in \mathcal{N}$, the ZF-based beamforming is designed as ${{\bf{w}}_k} = {\sqrt \alpha}{{\bf{h}}_{{\mathop{\rm AU}\nolimits} ,k}},\forall k \in \mathcal{K}$, where $\alpha$ represents the power control coefficient. Here ${{\bf{h}}_{{\mathop{\rm AU}\nolimits} ,k}}$ denotes the $k$th column of ${\bf{H}}_{{\mathop{\rm AU}\nolimits} }$, where ${{\bf{H}}_{{\mathop{\rm AU}\nolimits} }} = {{\bf{\bar G}}_{{\mathop{\rm AU}\nolimits} }}{\left( {{\bf{\bar G}}_{{\mathop{\rm AU}\nolimits} }^{\rm H}{{{\bf{\bar G}}}_{{\mathop{\rm AU}\nolimits} }}} \right)^{ - 1}}$ and ${{\bf{\bar G}}_{{\mathop{\rm AU}\nolimits} }} = \left[ {{{{\bf{\bar g}}}_{{\mathop{\rm AU}\nolimits} ,1}}, \cdots ,{{{\bf{\bar g}}}_{{\mathop{\rm AU}\nolimits} ,K}}} \right] \in {\mathbb{C}^{M \times N}}$, with ${\bf{\bar g}}_{{\mathop{\rm AU}\nolimits} ,k}^{\rm H}$ being the combined channel defined in \eqref{61}. Setting the transmit beamforming to ${{\bf{w}}_k} = {\sqrt \alpha}{{\bf{h}}_{{\mathop{\rm AU}\nolimits} ,k}}$, $\bf{P2}$ can be transformed to
\begin{align}
\left( \bf{P20} \right):\;\;\mathop {\max }\limits_{\alpha,{\theta _{ln}}} \;\;\;&\alpha \label{92}\\
{\rm s.t.}\;\;\,\,&\alpha {\left\| {{{{\bf{\bar h}}}_{{\mathop{\rm AU}\nolimits} ,m}}} \right\|^2} \le P,\;\;\,\,\forall m \in \mathcal{M},\label{93}\\
&{\theta _{ln}} \in \mathcal{F},\,\,\,\,\,\,\,\,\,\,\,\forall \,l \in \mathcal{L},\forall n \in \mathcal{N},\label{94}
\end{align}
where ${{{{\bf{\bar h}}}_{{\mathop{\rm AU}\nolimits} ,m}}}$ denotes the $m$th row of ${{\bf{H}}_{{\mathop{\rm AU}\nolimits} }}$.

Examining $\bf{P20}$ we can see that all users have the same SINR and the objective function $\alpha$ is closely related to ${\left\| {{{{\bf{\bar h}}}_{{\mathop{\rm AU}\nolimits} ,m}}} \right\|^2}$. If we want to maximize $\alpha$, we need to minimize the maximum of ${\left\| {{{{\bf{\bar h}}}_{{\mathop{\rm AU}\nolimits} ,m}}} \right\|^2},\forall m \in \mathcal{M}$. Therefore, $\bf{P20}$ can be recast to
\begin{align}
\left( \bf{P21}\right):\;\;\mathop {\min }\limits_{{\theta _{ln}}} \mathop {\max }\limits_{\forall m \in \mathcal{M}} \;&{\left\| {{{{\bf{\bar h}}}_{{\mathop{\rm AU}\nolimits} ,m}}} \right\|^2}\label{95}\\
{\rm s.t.}\;\;\;\;\;\;\,&{\theta _{ln}} \in \mathcal{F},\,\,\,\forall \,l \in \mathcal{L},\forall n \in \mathcal{N}.\label{96}
\end{align}

Since the channel matrix ${{\bf{H}}_{{\mathop{\rm AU}\nolimits} }}$ involves the inverse operation and the phase shift can only be selected from a discrete set, $\bf{P21}$ is a complicated nonlinear integer programming, which is still difficult to solve. Nevertheless, by fixing $LN-1$ phase shifts except for ${\theta _{ln}}$ in each iteration, we can find the suboptimal discrete solution of ${\theta _{ln}}$ via one-dimensional search over ${\theta _{ln}} \in \mathcal{F}$, i.e.,
\begin{equation}\label{97}
\theta _{ln}^ *  = \arg \;\mathop {\min }\limits_{{\theta _{ln}} \in \mathcal{F}} \;f\left( {{\theta _{ln}}} \right),
\end{equation}
where, for fixed discrete phase shifts ${\theta _{{l_1}{n_1}}},l_1 \neq l;n_1 \neq n$, the function $f\left( {{\theta _{ln}}} \right)$ is defined as
\begin{equation}\label{98}
f\left( {{\theta _{ln}}} \right) \buildrel \Delta \over = \mathop {\max }\limits_{\forall m \in \mathcal{M}} \;{\left\| {{{{\bf{\bar h}}}_{{\mathop{\rm AU}\nolimits} ,m}}} \right\|^2}.
\end{equation}

Based on \eqref{97}, we can alternately optimize each of the $LN$ phase shifts in an iterative manner by fixing the other $LN-1$ phase shifts. Until all $f\left( {{\theta _{ln}}} \right)$ are not decreasing and achieve convergence, suboptimal discrete phase shifts can be determined for all RISs. Based on these suboptimal discrete phase shifts, the corresponding ZF-based beamforming ${{\bf{w}}_k} = {\sqrt \alpha}{{\bf{h}}_{{\mathop{\rm AU}\nolimits} ,k}}$ and the power control coefficient $\alpha  = \frac{P}{{\mathop {\max }\limits_{m \in \mathcal{M}} \;{{\left\| {{{{\bf{\bar h}}}_{{\mathop{\rm AU}\nolimits} ,m}}} \right\|}^2}}}$ can be determined accordingly. The ZF-based successive refinement algorithm for solving $\bf{P2}$ is summarized in ${\bf{Algorithm\:6}}$.

\begin{algorithm}\label{alg:algo6}
\caption{ZF-based successive refinement algorithm for $\bf{P2}$}
\begin{algorithmic}[1]
\STATE {Input the channels ${\bf{g}}_{{\mathop{\rm RU}\nolimits} ,lk}^{\rm H}$,${{\bf{G}}_{{\mathop{\rm AR}\nolimits} ,l}}$, ${{\bf{g}}_{{\mathop{\rm AU}\nolimits} ,k}}$ and the initial phase shifts ${\theta _{ln}}$. Then, set the beamforming matrix ${{\bf{H}}_{{\mathop{\rm AU}\nolimits} }}$, the initial iteration number $t=1$ and the maximum number of iterations $T=300$.}
\STATE {\textbf{Repeat:}}
\STATE {With all phase shifts, except for ${\theta _{ln}}$, are fixed, set $\theta _{ln}^*  =\arg \;\mathop {\min }\limits_{{\theta _{ln}} \in \mathcal{F}} \;f\left( {{\theta _{ln}}} \right)$ as the suboptimal phase shift, and update $\theta _{ln}=\theta _{ln}^*$ and ${{\bf{H}}_{{\mathop{\rm AU}\nolimits} }}$.}
\STATE {Set $t=t+1$ and choose a different phase shift as the optimization variable in the next iteration.}
\STATE {\textbf{Until}\:$t=T$ or all $f\left( {{\theta _{ln}}} \right)$ are not decreasing and reach converge. Denote ${\theta _{ln}},\forall \,l \in \mathcal{L},\forall n \in \mathcal{N}$ as suboptimal phase shifts and go to Step 6.}
\STATE {Based on the suboptimal phase shifts, obtain ${{\bf{w}}_k} = {\sqrt \alpha}{{\bf{h}}_{{\mathop{\rm AU}\nolimits} ,k}},\forall \,k \in \mathcal{K}$ as the optimal transmit beamforming, where $\alpha  = \frac{P}{{\mathop {\max }\limits_{m \in \mathcal{M}} \;{{\left\| {{{{\bf{\bar h}}}_{{\mathop{\rm AU}\nolimits} ,m}}} \right\|}^2}}}$.}
\end{algorithmic}
\end{algorithm}

According to ${\bf{Algorithm\:6}}$, the computational complexity is mainly determined by Step 3, which involves the computational complexity of $\mathcal{O}\left( {{{\hat I}_{{\rm{iter}}}}B} \right)$, where ${\hat I}_{{\rm{iter}}}$ is the number of iterations for the convergence of ${\bf{Algorithm\:6}}$.

\section{Numerical Results}
In this section, simulation results are provided to validate the effectiveness of our proposed algorithms for RIS-aided cell-free massive MIMO systems.

\subsection{Simulation Parameters}
\begin{figure}[thb!]
\centering
\includegraphics[scale=0.5]{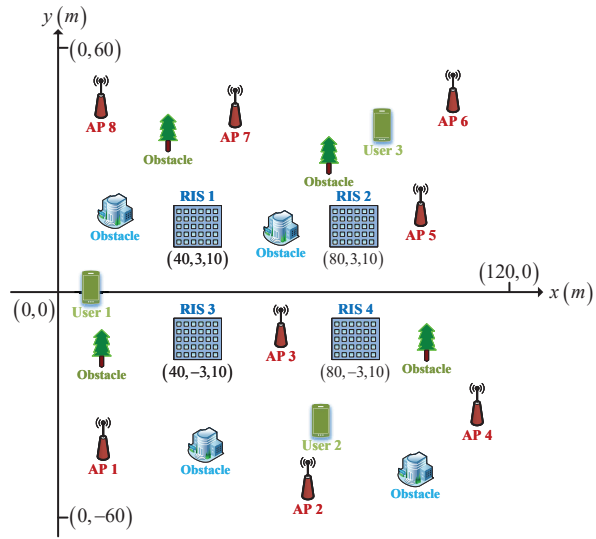}
\caption{Simulation scenario where eight APs provide services to three users with the assistance of four RISs.}\label{fig3}
\end{figure}

Without loss of generality, a three-dimensional (3D) coordinate system is considered in Fig. \ref{fig3}. In this system, eight single-antenna APs and four RISs provide services to three single-antenna users within a square area of 120 m $\times $120 m, where the APs with a height of 5 m and the users with a height of 1.65 m are randomly distributed, and four RISs with a height of 10 m are located at (40 m, 3 m, 10 m), (80 m, 3 m, 10 m), (40 m, $-$3 m, 10 m) and (80 m, $-$3 m, 10 m), respectively. All the RISs are configured with uniform rectangular arrays, each having $N$ reflecting elements. For the channel model involved in this section, we consider similar settings as those in \cite{Wu19}. The distance-dependent path loss can be modeled as $\beta  = {c_0}{\left( {\frac{{{d_1}}}{{{d_0}}}} \right)^{ - \alpha }}$, where ${c_0}=-30$ dB denotes the path loss at the reference distance ${d_0}=1$ m, ${d_1} \in \left\{ {{d_{{\mathop{\rm AR}\nolimits} ,ml}},{d_{{\mathop{\rm RU}\nolimits} ,lk}},{d_{{\mathop{\rm AU}\nolimits} ,mk}}} \right\}$ and $\alpha  \in \left\{ {{\alpha _{{\mathop{\rm AR}\nolimits} ,ml}},{\alpha _{{\mathop{\rm RU}\nolimits} ,lk}},{\alpha _{{\mathop{\rm AU}\nolimits} ,mk}}} \right\}$ denote the distance and the path loss exponent between AP-RIS link, RIS-user link and AP-user link, respectively. To account for small-scale fading, we assume that all channels are Rician fading. Take an AP-RIS link as an example, the channel between the $m$th AP and the $l$th RIS is given by
\begin{equation}\label{99}
{{\bf{g}}_{{\mathop{\rm AR}\nolimits} ,ml}} = \sqrt {\frac{{{\beta _{{\mathop{\rm AR}\nolimits} ,ml}}{\kappa _{{\mathop{\rm AR}\nolimits} ,ml}}}}{{1 + {\kappa _{{\mathop{\rm AR}\nolimits} ,ml}}}}} {{\bf{\bar h}}_{{\mathop{\rm AR}\nolimits} ,ml}}+\sqrt {\frac{{{\beta _{{\mathop{\rm AR}\nolimits} ,ml}}}}{{1 + {\kappa _{{\mathop{\rm AR}\nolimits} ,ml}}}}} {{\bf{h}}_{{\mathop{\rm AR}\nolimits} ,ml}},
\end{equation}
where $\beta _{{\mathop{\rm AR}\nolimits},ml}$, ${{\bf{\bar h}}_{{\mathop{\rm AR}\nolimits} ,ml}}\in {\mathbb{C}^{N \times 1}}$, ${{\bf{h}}_{{\mathop{\rm AR}\nolimits} ,ml}}\in {\mathbb{C}^{N \times 1}}$ and ${\kappa _{{\mathop{\rm AR}\nolimits} ,ml}}$ denote the distance-dependent path loss, the deterministic line-of-sight (LOS) component, the Rayleigh fading component and the Rician factor, respectively. The channel between the $l$th RIS and the $k$th user, ${{\bf{g}}_{{\rm RU},lk}^{\rm H}}$, is also generated in a similar manner. Different from ${{\bf{g}}_{{\mathop{\rm AR}\nolimits} ,ml}}$ and ${{\bf{g}}_{{\rm RU},lk}^{\rm H}}$, since APs and users are not placed at high positions, we assume that the direct link between the $m$th AP and the $k$th user, namely ${{g}_{{\rm AU},mk}}$, has a probability $p_{\rm AU}$ of being blocked by obstacles. Therefore,  the distance-dependent path loss between the $m$th AP and the $k$th user is expressed as ${\beta _{{\mathop{\rm AU}\nolimits} ,mk}} = {\bar \beta _{{\mathop{\rm AU}\nolimits} ,mk}}{\rho _{{\mathop{\rm AU}\nolimits} ,mk}}$, where ${\bar \beta _{{\mathop{\rm AU}\nolimits} ,mk}}$ is generated in a similar manner as ${{\beta}_{{\mathop{\rm AR}\nolimits} ,ml}}$. The binary variable ${\rho _{{\mathop{\rm AU}\nolimits} ,mk}}$ is 0 with a probability of $p_{\rm AU}$ and 1 with a probability of $1 - p_{\rm AU}$.

Besides, the transmit power at each AP, the noise power, and the normalized SNR are set as $\bar P=0$ dBm, $N_0=-80$ dBm, and $P = {{\bar P} \mathord{\left/{\vphantom {{\bar P} {{N_0}}}} \right.\kern-\nulldelimiterspace} {{N_0}}}$, respectively. Other system parameters are set as follows: $p_{\rm AU}=0.2$, ${\alpha _{{\mathop{\rm AR}\nolimits} ,ml}}{\rm{ = }}1$, ${\alpha _{{\mathop{\rm RU}\nolimits} ,lk}}{\rm{ = }}1.5$, ${\alpha _{{\mathop{\rm AU}\nolimits} ,mk}}{\rm{ = }}3.5$, ${\kappa _{{\mathop{\rm AR}\nolimits} ,ml}}{\rm{ = }}10$ dB, ${\kappa _{{\mathop{\rm RU}\nolimits} ,lk}}{\rm{ = }}5$ dB and ${\kappa _{{\mathop{\rm AU}\nolimits} ,mk}}{\rm{ = 0}}$. In this section, all numerical results are averaged over 100 independent channel realizations. For comparison with the proposed algorithms, we study two benchmark schemes:

\begin{enumerate}

\item \emph{Random phase shifts:} First, all phase shifts at each RIS are randomly selected in ${\theta _{ln}} \in \left[ {0,2\pi } \right)$ for the case of continuous phase shifts, or ${\theta _{ln}} \in \mathcal{F}$ for the case of discrete phase shifts. Then, based on the chosen phase shifts, MRT beamforming is performed at each AP to maximize the achievable rate for the single-user setup. For the multi-user setup, with the chosen phase shifts, we use ${\bf{Algorithm\:3}}$ to obtain the transmit beamforming that maximizes the minimum achievable rate among all users.

\item \emph{Without RISs:} This is the conventional cell-free massive MIMO system without deploying RISs. Based on the AP-user direct link, we perform MRT at each AP for the single-user setup, and apply ${\bf{Algorithm\:3}}$ at each AP for the multiuser-user setup.

\end{enumerate}

\subsection{Results for the Single-User System}
\vspace{-0.4cm}
\begin{figure}[thb!]
\setlength{\abovecaptionskip}{0.cm}
\setlength{\belowcaptionskip}{-0.cm}
\centering
\includegraphics[scale=0.425]{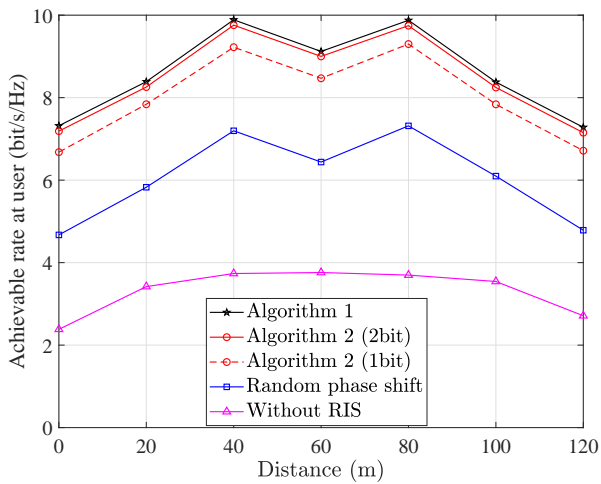}
\caption{Achievable rate versus the distance between the user and the origin.}\label{fig4}
\end{figure}

We first consider the special single-user scenario. For Fig. \ref{fig4}, we study a single-user system with $N=12$, in which a single user moves in a straight line from (0 m, 0 m) to (120 m, 0 m). Compared to the scheme without using RISs, we can observe two obvious peaks when the user is located at (40 m, 0 m) and (80 m, 0 m) for the schemes with ${\bf{Algorithm\:1}}$, ${\bf{Algorithm\:2}}$ and random phase shift, respectively. This phenomenon shows that when the user is close to a RIS, the received signal strength at the user can be significantly enhanced by using the reflected signals from the RIS. This also shows that deploying RISs can improve the user rate performance and signal coverage. Then, compared to the scheme of random phase shifts, both ${\bf{Algorithm\:1}}$ for continuous phase shifts and ${\bf{Algorithm\:2}}$ for discrete phase shifts can significantly increase the achievable rate of the user. This verifies the effectiveness of the proposed algorithms for the single-user system. We can also see that the performance of ${\bf{Algorithm\:2}}$ with 2-bit phase resolution can practically approximate the performance of ${\bf{Algorithm\:1}}$. This shows that the joint optimization algorithm for discrete phase shifts is very suitable and attractive for practical applications in real scenarios.

\vspace{-0.3cm}
\begin{figure}[thb!]
\setlength{\abovecaptionskip}{0.cm}
\setlength{\belowcaptionskip}{-0.cm}
\centering
\includegraphics[scale=0.425]{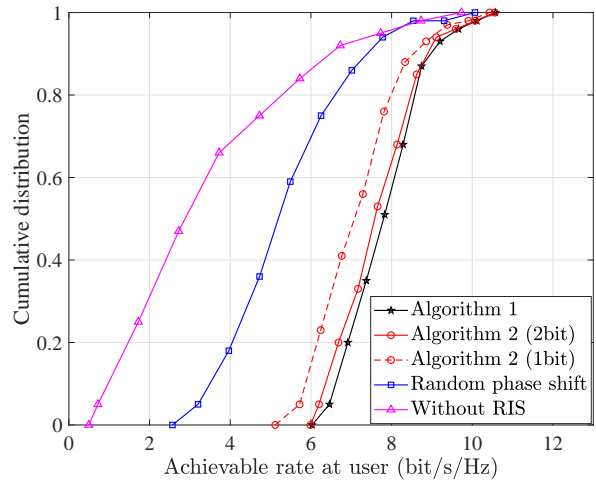}
\caption{Cumulative distribution of the achievable rate.}\label{fig5}
\end{figure}


Different from the single-user scenario considered for Fig. \ref{fig4} in which the user  only moves along a straight line, we consider the scenario that the user can move to any position and plot in Fig. \ref{fig5} the cumulative distribution of the achievable rate with $N=12$. We can see that the $95\%-{\rm{likely}}$ achievable rate obtained by \textbf{Algorithm 1}, \textbf{Algorithm 2} with 2-bit resolution and \textbf{Algorithm 2} with 1-bit resolution are about 6.45 bit/s/Hz, 6.19 bit/s/Hz, and 5.71 bit/s/Hz, which are 2.01, 1.93, and 1.78 times higher than that achieved with the scheme based on random phase shifts (about 3.2 bit/s/Hz), respectively. The simulation results also verifies that \textbf{Algorithm 1} and \textbf{Algorithm 2} can significantly increase the achievable rate of the user. In addition, for the $95\%-{\rm{likely}}$ achievable rate, \textbf{Algorithm 2} with 1-bit or 2-bit resolution can reach $88{\%}$ or $95{\%}$ of the performance of \textbf{Algorithm 1}, respectively. Therefore, for practical applications, \textbf{Algorithm 2} with 2-bit phase resolution would be the most attractive.
\begin{figure}[thb!]
\setlength{\abovecaptionskip}{0.cm}
\setlength{\belowcaptionskip}{-0.cm}
\centering
\includegraphics[scale=0.425]{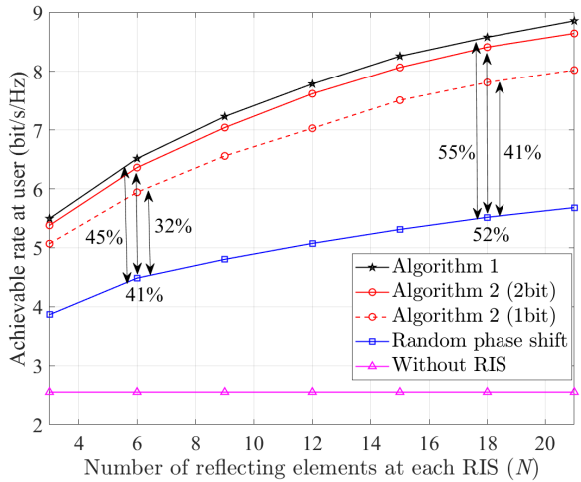}
\caption{Achievable rate versus the number of reflecting elements.}\label{fig6}
\end{figure}

Fig. \ref{fig6} presents the user's achievable rate versus the number of reflecting elements at each RIS ($N$). First, the results show that as $N$ increases, the user's achievable rate also increases. Then, it can be observed that with $N$ increasing, performance gaps between \textbf{Algorithm 1} and \textbf{Algorithm 2} and the benchmark scheme of random phase shifts gradually widen. For example, compared with the case of $N=6$, the gains by the proposed algorithms in the case of $N=18$ increase by about $10{\%}$. This shows that as $N$ increases, proper designs of the phase shifts at RISs play a more important role. Finally, we also see that the proposed algorithms considerably outperform the two benchmark schemes. This shows the benefit of our proposed designs.

\subsection{Multi-User System}
\begin{figure}[thb!]
\setlength{\abovecaptionskip}{0.cm}
\setlength{\belowcaptionskip}{-0.cm}
\centering
\includegraphics[scale=0.425]{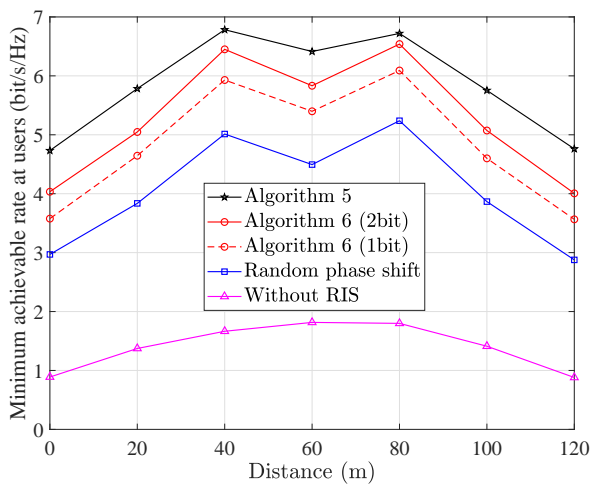}
\caption{Minimum achievable rate versus the distance between the users and the origin.}\label{fig7}
\end{figure}

Next, we consider a general multi-user scenario with three users. When three users are in a circle of radius 1 and move from (0 m, 0 m) to (120 m, 0 m) in a straight line, Fig. \ref{fig7} shows the minimum achievable rate among users versus the distance between the users and the origin. Here, we assume $N=12$. Similar to the phenomenon in Fig. \ref{fig4}, we can observe that when the users are close to a RIS, the minimum achievable rate reaches two peaks for the schemes with ${\bf{Algorithm\:5}}$, ${\bf{Algorithm\:6}}$ and random phase shifts. In addition, compared to the two benchmark schemes, both ${\bf{Algorithm\:5}}$ and ${\bf{Algorithm\:6}}$ can effectively increase the minimum achievable rate when the users are in different positions. These phenomena show the benefits of using RISs and the proposed algorithms for the multi-user system.

\begin{figure}[thb!]
\setlength{\abovecaptionskip}{0.cm}
\setlength{\belowcaptionskip}{-0.cm}
\centering
\includegraphics[scale=0.425]{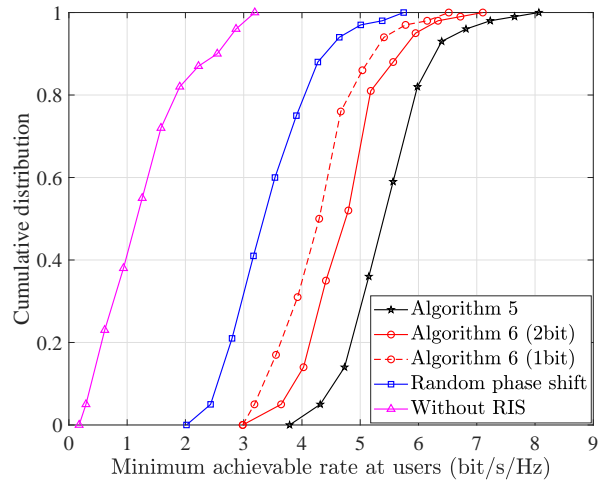}
\caption{Cumulative distribution of the minimum achievable rate.}\label{fig8}
\end{figure}

When three users are randomly distributed in a coverage area, Fig. \ref{fig8} studies the cumulative distribution of the minimum achievable rate in the case of $N=12$. Compared to the scheme with random phase shifts (about 2.43 bit/s/Hz), the $95\%-{\rm{likely}}$ minimum achievable rates obtained by \textbf{Algorithm 5} (about 4.31 bit/s/Hz), \textbf{Algorithm 6} with 2-bit resolution (about 3.64 bit/s/Hz) and \textbf{Algorithm 6} with 1-bit resolution (about 3.18 bit/s/Hz) increase roughly by $77{\%}$, $49{\%}$ and $30{\%}$, respectively. The results thus clearly show the effectiveness of \textbf{Algorithm 5} and \textbf{Algorithm 6} in improving the minimum achievable rate as compared to the two benchmark schemes. Besides, in view of the fact that \textbf{Algorithm 6} with 2-bit resolution can basically achieve $85{\%}$ of the performance of \textbf{Algorithm 5}, the discrete phase shift optimization algorithm for the multi-user system is very attractive for practical scenarios.
\vspace{-0.4cm}
\begin{figure}[thb!]
\setlength{\abovecaptionskip}{0.cm}
\setlength{\belowcaptionskip}{-0.cm}
\centering
\includegraphics[scale=0.425]{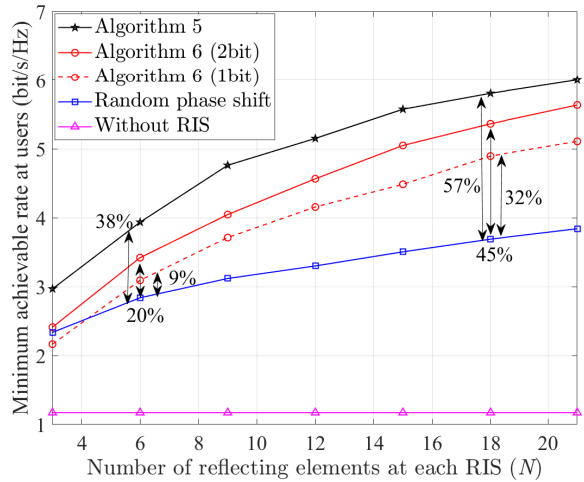}
\caption{Minimum achievable rate versus the number of reflecting elements.}\label{fig9}
\end{figure}

In Fig. \ref{fig9}, we present the minimum achievable rate versus the number of reflecting elements at each RIS ($N$). Similar observations are made as in Fig. \ref{fig6}: When $N$ increases from 6 to 18, using the scheme of random phase shifts as a benchmark, \textbf{Algorithm 5} or \textbf{Algorithm 6} can improve the rate performance by approximately $19{\%}$ and $24{\%}$, respectively. The simulation results show that as the number of reflecting elements at each RIS increases, both \textbf{Algorithm 5} and \textbf{Algorithm 6} can stronger exert their superiority in the multi-user system.

\begin{figure}[thb!]
\setlength{\abovecaptionskip}{0.cm}
\setlength{\belowcaptionskip}{-0.cm}
\centering
\includegraphics[scale=0.425]{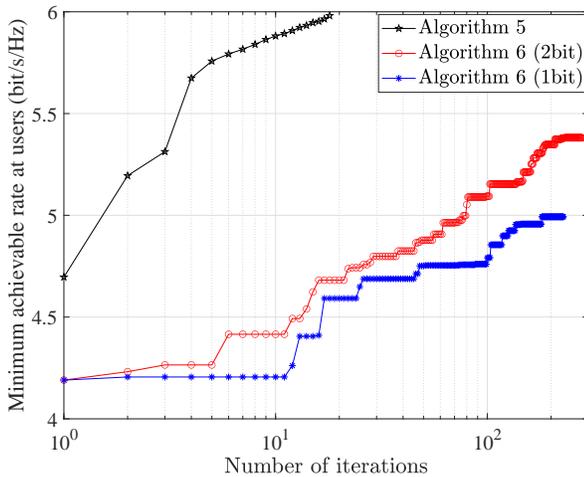}
\caption{Minimum achievable rate versus the number of iterations.}\label{fig10}
\end{figure}

Fig. \ref{fig10} shows the convergence behavior of \textbf{Algorithm 5} and \textbf{Algorithm 6}, where we set $N=12$. First, it is observed that \textbf{Algorithm 5} is terminated at 18 iterations, while \textbf{Algorithm 6} with 1-bit and 2-bit resolution can converge on 230 and 280 iterations, respectively. The reason that \textbf{Algorithm 5} terminates prematurely is because the SDR results in a reduced minimum SINR during the iteration due to Gaussian randomization. Although the iteration of \textbf{Algorithm 5} was quickly terminated, its performance is still higher than that of \textbf{Algorithm 6}. In addition, the performance of \textbf{Algorithm 6} with 2-bit resolution is better than that of 1-bit resolution. This shows that as the number of quantization bits increases, \textbf{Algorithm 6} can yield a better performance advantage.

\vspace{-0.4cm}
\vspace{-0.4cm}

\section{Conclusions}
In this paper, we developed joint optimization algorithm for RIS-aided cell-free massive MIMO systems. Specifically, the continuous transmit beamforming at the APs and the continuous/discrete phase shifts at RISs were jointly optimized to maximize the minimum achievable rate among users, while meeting transmit power constraints at different APs. To tackle the max-min fairness problem, for the case of continuous phase shifts, we proposed a SDP-based algorithm (Algorithm 1) for the single-user setup, and an alternating optimization algorithm based on SDP and SCOP (Algorithm 5) for the multi-user setup. Besides, for the more practical case of discrete phase shifts, we proposed an ILP-based algorithm (Algorithm 2) for the single-user scenario and a ZF-based successive refinement algorithm (Algorithm 6) for the multi-user scenario. Simulation results revealed that, with the assistance of low-cost and energy-saving RISs, the proposed joint optimization algorithms can guarantee the users' QoS in different geographical locations compared with the benchmark schemes of random phase shifts and without the RISs. An interesting direction for a future work is an extension of our proposed algorithms to other key performance metrics (e.g. energy efficiency or resource efficiency).

\balance

\end{document}